\documentclass[fleqn,usenatbib]{mnras}

\usepackage{newtxtext,newtxmath}
 
\usepackage[T1]{fontenc}

\DeclareRobustCommand{\VAN}[3]{#2}
\let\VANthebibliography\thebibliography
\def\thebibliography{\DeclareRobustCommand{\VAN}[3]{##3}\VANthebibliography}

\usepackage{graphicx}	
\usepackage{amsmath}	
\usepackage{amssymb}	
\usepackage{cancel}

\title[HPC W-stacking]{High Performance W-stacking for Imaging Radio Astronomy Data: a Parallel and Accelerated Solution}

\author[]{
Claudio Gheller,$^{1}$\thanks{E-mail: claudio.gheller@gmail.com}
Giuliano Taffoni$^{2}$
and David Goz,$^{2}$
\\
$^{1}$Institute of Radioastronomy, INAF, Via Gobetti 101, 40121 Bologna, Italy\\
$^{2}$Astronomical Observatory of Trieste INAF, via GB Tiepolo 11, 34143 Trieste, Italy
}

\date{Accepted XXX. Received YYY; in original form ZZZ}

\pubyear{2015}

\begin{document}
\label{firstpage}
\pagerange{\pageref{firstpage}--\pageref{lastpage}}
\maketitle

\begin{abstract}
Current and upcoming radio-interferometers are expected to produce volumes of data of increasing size that need to be processed in order to generate the corresponding sky brightness distributions through imaging. This represents an outstanding computational challenge, especially when large fields of view and/or high resolution observations are processed. We have investigated the adoption of modern High Performance Computing systems specifically addressing the gridding, FFT-transform and w-correction of imaging, combining parallel and accelerated solutions. We have demonstrated that the code we have developed can support dataset and images of any size compatible with the available hardware, efficiently scaling up to thousands of cores or hundreds of GPUs, keeping the time to solution below one hour even when images of the size of the order of billion or tens of billion of pixels are generated. In addition, portability has been targeted as a primary objective, both in terms of usability on different computing platforms and in terms of performance. The presented results have been obtained on two different state-of-the-art High Performance Computing architectures. 
\end{abstract}


\section{Introduction}
\label{sec:intro}

In the next decade, current and upcoming radio-interferometers, like the Low Frequency Array \citep[LOFAR,][]{2013A&A...556A...2V}, MeerKAT\citep[][]{2016mks..confE...1J}, the Murchison Widefield Array\citep[MWA,][]{2010rfim.workE..16M}, the Australian Square Kilometre Array Pathfinder \citep[ASKAP,][]{2007PASA...24..174J}, in the perspective of the Square Kilometre Array (SKA MID and LOW)\footnote{https://www.skatelescope.org/}, will produce huge volumes of data. This will represent not only an invaluable opportunity for scientists, but also an outstanding technological challenge. The expected data volume will be hard to manage with traditional approaches. Data will have to be stored in dedicated facilities, providing the necessary capacity at the highest performance. Corresponding data processing will have to be performed local to the data, exploiting available High Performance Computing (HPC) resources. Data reduction and imaging software tools will have to be adapted, if not completely re-designed, in order to efficiently run at scale.

In modern HPC systems, performance is being achieved through many-core and accelerated computing (based, for instance, on GPUs), and current trends suggest that some form of heterogeneous computing will continue to be prevalent in emerging architectures.
Therefore, the ability to fully exploit new heterogeneous and many-core solutions is of paramount importance towards achieving optimal performance. On the other hand, with the increasing size and complexity of observational data, it is of primary importance for scientists to be able to exploit all available hardware in emerging HPC environments.
Exploiting these novel hybrid architectures is non-trivial however, due to the challenges presented by mixed hardware computing and the increasing levels of architectural parallelism. New algorithms and numerical and computational solutions are required.
\smallskip

In radio interferometry, data processing is generally performed in various steps (often repeated iteratively), based on the work of \cite{1996A&AS..117..137H, 2009IEEEP..97.1472R, 2011A&A...527A.106S, 2011A&A...527A.107S}. Firstly, the calibration step aims at estimating and correcting for time, frequency, antenna and direction dependent instrumental errors \citep{2008A&A...487..419B, 2008ISTSP...2..647C, 2013A&A...553A.105T, 2013ApJ...770...91B, 2017AJ....154...56J, 2017AJ....154..197B}. Various software tools address calibration, like DPPP \citep{2018ascl.soft04003V}, CUBICAL \citep{kenyon2018} or KillMS\footnote{https://github.com/saopicc/killMS}.
This is followed by {\it Imaging}, that is the processes of Fourier-transforming the calibrated visibilities into images ( for example DDFacet, addressing anyway also calibration \citep{2018A&A...611A..87T}, and WSClean, \citep{offringa-wsclean-2014}. Then deconvolution \citep[see][]{cornwell:1999ASPC..180.....T} corrects the resulting images for the incomplete sampling of the Fourier plane. Deconvolution is implemented in terms of the consolidated and widely used Clean algorithm \citep{1974AAS...15..417H} in its different variants \citep[][among the most widely used]{1980AA....89..377C, 1984AJ.....89.1076S}, or other approaches, like the Maximum Entropy Model \citep{1985A&A...143...77C}, the MORESANE model \citep{2015AA...576A...7D}, the Sparsity Averaging Reweighted Analysis (SARA) family of deconvolution algorithms defined in the context of optimisation theory, and implemented in the Puri-PSI library\footnote{https://basp-group.github.io/Puri-Psi/}, have been demonstrated to bring joint precision and scalability of the reconstruction process
 \citep{Carrillo_2012, purify, Onose_2016, 2018MNRAS.473.1038P,Thouvenin2020,Thouvenin2022}, 
also including a joint approach for calibration and deconvolution \citep{Repetti2017,Dabbech_2011},
 the SASIR method \citep{2015JInst..10C8013G}, the RESOLVE Bayesian method \citep{2016AA...586A..76J} and the clean multiscale deconvolution approach \citep{2011A&A...532A..71R, 0004-637X-635-1-73}. Finally, denoising \citep[we refer to][for a comprehensive review]{2020A&A...643A..43R} and source detection and characterisation are performed using tools that are usually available within source finding software packages, like PyBDSF \citep{2021A&A...645A..89M}, SoFiA \citep{2015MNRAS.448.1922S} and AEGEAN \citep{2012MNRAS.422.1812H} among the most up-to-date. Also of note are comprehensive and widely adopted software platforms, able to perform most of the previous tasks, like CASA \citep{2007ASPC..376..127M}, AIPS \citep{Greisen2003}, MIRIAD \citep{1995ASPC...77..433S} and ASKAPsoft \citep{2020ASPC..527..591W}.  Finally, it is worth mentioning the two first applications of Deep Learning for imaging presented by \cite{Terris_2022} and \cite{2022ApJ...939L...4D}.
 
Imaging represents one of the most computational demanding steps of the processing pipeline, both in terms of memory request and in terms of computing time, associated to operations like gridding, i.e. the convolutional resampling of the observed data on a computational mesh, or FFT transforming, to move from Fourier to real space and vice-versa. The computational requirements increase further when observations with large fields of view (as typically happens with current radio interferometers and at the lowest frequencies) are considered, since curvature effects cannot be neglected and the problem becomes fully three dimensional, with the introduction of the {\it w-term} correction (see Section \ref{w-stacking}, \cite{2008ISTSP...2..647C,offringa-wsclean-2014}).

Various algorithms to cope with such three dimensional problem have been implemented, several of them being enabled and optimized to exploit specific computational resources (eg. GPUs or multi-core platforms). Currently, the most widely adopted approach is the $w$-projection algorithm
\citep{offringa-wsclean-2014}, 
in which the $w$-term is expressed as a convolution in Fourier space and it is applied directly to the visibilities. Here $w$ represents the third direction in the ($u,v,w$) coordinates system of the visibility space (see Section \ref{w-stacking} for details). The $w$-projection algorithm has been parallelized and ported to the GPU by \cite{2019arXiv190503213L}, who however limited the data distribution to the visibility data. A CUDA GPU based implementation is proposed also by \cite{2014arXiv1403.4209M}. An alternative approach is represented by the $w$-stacking technique \citep{2008ISTSP...2..647C}, in which the data is partitioned in $w$, and the $w$ correction is applied in image space by multiplying each $w$ plane after FFT transform. This solution has been adopted by WSClean \citep{offringa-wsclean-2014}, 
followed by the work of \cite{arras21} and \cite{Ye_2022}, optimising the original implementation. However, only multi-core, shared memory architectures are supported. The WSClean software implements also the IDG solution supporting both multicore and GPU implementations \citep{2018A&A...616A..27V,2020A&C....3200386V}. The gridding step has also been enabled to the GPU by \cite{Merry2016FasterGC}. The $w$-stacking and $w$-projection approaches have been combined and parallelized on distributed memory systems by \cite{2020PASA...37...41P}. A further solution is represented by Faceting, in which the $w$ correction is taken to be constant or linear over small regions of the sky \citep{1992A&A...261..353C}. The image is constructed by piecing together many different facets. For this approach, parallel multi-core support can be found in the DDFacet package \citep{2018A&A...611A..87T}. 

The aforementioned solutions are only partially capable of fully facing the challenge related to big data exploiting modern, hybrid HPC solutions. In particular, parallel computing is not effectively used to process increasingly larger datasets and images, that cannot fit a single memory systems, in conjunction with accelerated architectures that can dramatically reduce the time to solution. 

In this work, we have addressed such challenge focusing on those steps of the imaging pipeline that characterise the $w$-stacking algorithm, namely gridding, FFT-transform and w-correction. Hereafter, we refer to this combination of steps as {\it w-stacking gridder}. These steps are suitable to distributed memory parallelism, exploiting parallel FFT solutions and relying on a Cartesian 3D computational mesh, that can be easily distributed and efficiently managed across different processing units, potentially leading to a good scalability on large HPC architectures.

Additionally, we have targeted portability as one of our prime goals. Given the large number of different HPC solutions currently available, and since the development of applications codes spans a period of time longer than the typical timescale of HPC architecture evolution, our objective is to design a code easily portable on different computing platforms both in terms of code usability and in terms of performance. In this way, scientists can effectively exploit all the available supercomputing resources, without finding major obstacles in building and running the code on a new system, obtaining an acceptable performance and scalability anywhere. 

The code has been developed adopting the C programming language standard (with extensions to C++ only to support GPUs through CUDA) and it has not been fine tuned to any specific computational architecture, in order to avoid introducing specialised solutions that would limit its portability. In order to have a portable solution even for the GPU implementation, alongside CUDA, we have experimented also the OpenMP support to offload operations to accelerators. We have not adopted performance portable solutions like OpenCL or Kokkos, since they introduce unwanted dependencies from additional libraries and applications.

We present the results obtained on two state-of-the-art supercomputing platforms: The Juwels Booster Module at the J\"ulich Supercompting Centre (Germany, hereafter {\it Juwels}) and Marconi 100 at the CINECA Supercomputing Centre (Italy, hereafter {\it M100}), that are ranked as third and sixth HPC systems in Europe in the TOP500 list\footnote{https://www.top500.org} of June 2022.  
Compilation on the two different platforms and with different compilers is done with a single source code version. The only required customisation is related to compilation flags and environmental variables. All tests have been performed using LOFAR datasets, representative of current radio-observations.

The paper is organised as follows. The methods used for performing the w-stacking is described in Section 2. In Section 3 we describe the different HPC solutions adopted for the code. In Section 4 the results of the performance and scalability tests are presented and discussed. Conclusions are drawn in Section 5.

\section{The $W$-stacking Gridder}
\label{w-stacking}

An interferometer measures complex visibilities $V$ related to the sky brightness distribution as:
\begin{equation}
\begin{split}
V(u,v,w) = &\int\int \frac{I(l,m)}{\sqrt{1-l^2-m^2}} \times \\ 
         &e^{-2\pi i \left(ul + vm + w(\sqrt{1-l^2-m^2}-1)\right)} dl dm,
\label{eq:visI}
\end{split}
\end{equation}

where $u,v,w$ is a baseline coordinate in the coordinate system of the antennas and $I$ is the spectral brightness and $l,m$ are the sky coordinates. For small fields of view, the term $\sqrt{1-l^2-m^2}$ is close to one, and Equation \ref{eq:visI} is an ordinary two-dimensional Fourier transform, which, in order to speed-up the computation, is solved by using a Fast Fourier Transforms (FFT) based approach. This however, requires to map visibilities, that are point like data, to a regular mesh, which discretizes the ($u$,$v$) space. This is accomplished by convolving the visibility data with a finite-size kernel, which converts it to a continuous function, which can then be FFT transformed.

When large Fields of View (FoV) are observed at once, visibility data from non-coplanar interferometric radio telescopes cannot be accurately imaged with a two-dimensional Fourier transform: the imaging algorithm needs to account for the {\it $w$-term}. This term describes the deviation of the array from a plane. 

A possible approach to account for the $w$-term is represented by the $w$-stacking method \citep{2008ISTSP...2..647C}, in which the computational mesh has a third dimension in the $w$ direction and visibilities are mapped to the closest $w$-plane. Once gridding is completed, each w-plane is Fourier transformed separately, and a phase correction is applied as:
\begin{equation}
\begin{split}
\frac{I(l,m)\left(w_{\max} - w_{\min}\right)}{\sqrt{1-l^2-m^2}} = &\int\limits_{w_{\min}}^{w_{\max}} e^{2\pi i w(\sqrt{1-l^2-m^2}-1)} \times 
\\
&\iint V(u,v,w)  e^{2\pi i \left(ul + vm\right)} du dv dw.
\label{eq:wstacking}
\end{split}
\end{equation}
For all the details, we refer to \cite{offringa-wsclean-2014}.\\

\begin{figure*}
\centering
\includegraphics[width=0.50\textwidth]{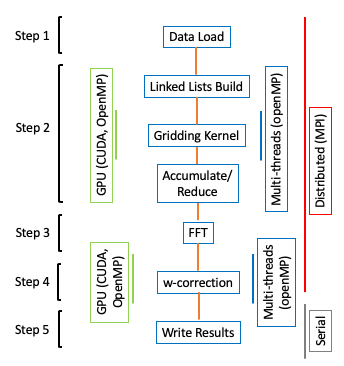}
\caption{Schematic code architecture. Different kind of HPC enabling are highlighted.}
\label{fig:workflow}
\end{figure*}

The $w$-stacking methodology has been implemented in a code  using the C programming language, with some C++ extensions required by the CUDA GPU implementation, adopting a procedural programming approach. Schematically, the resulting code is shown in Figure \ref{fig:workflow}, in which we highlight the parts that have been subject to a distributed parallel implementation and those that have been also accelerated.

Two main data structures characterise the algorithm. The first is an unstructured dataset storing the ($u,v,w$) coordinates of the antennas array baselines at each measurement time. Each baseline has a number of associated visibilities, which is determined by the frequency bandwidth, the frequency resolution and the number of polarisations. A further quantity, the weight, is also assigned to each measurement and polarisation. The second, is a Cartesian computational mesh of size $N_u \times N_v \times N_w$, where $N_u$, $N_v$ and $N_w$ are the number of cells in the three coordinate directions (see Figure \ref{fig:cube}). The convolved visibilities and their FFT transformed counterpart are calculated on the mesh. The two data structures determine the memory request of the algorithm. They are evenly distributed among different computing units as described in Section \ref{data_dist}.

The code consists of five main algorithmic components, each supporting different types of HPC implementations. The first component takes care of reading observational data from binary files stored on the disk. 
The files are read in parallel, assigning the same fraction of the data to each different parallel task (MPI task, see Section \ref{data_dist}).

The following component performs the gridding of the visibilities. Gridding is done in successive rectangular slabs along the v-axis
The gridding procedure consists in three sub-steps. In the first sub-step a linked list is created for each slab, concatenating the data with u-v coordinates inside the corresponding slab. 
The second sub-step is represented by the gridding kernel. The linked lists is traversed selecting the visibilities belonging to a given slab to convolve through a gaussian kernel:
\begin{equation}
    \tilde V(u_i,v_j,w_k) = \sum_{m \in {\rm measures}} V_m G((u_m,v_m,w_m),(u_i,v_j,w_k)),
\label{convolve}
\end{equation}
where $m$ id the m-th measurement, $(u_m,v_m,w_m)$ are its coordinates, $(u_i,v_j,w_k)$ is a computational grid point, $V_m$ is the measured visibility and $\tilde V$ is the visibility convolved on the mesh. The $G$ kernel is a Gaussian convolution function with a kernel size of 7 cells, following \cite{offringa-wsclean-2014}. Although Gaussian kernels are not used by any production imagers, leading to a high aliasing response in image space, in this work we have chosen to adopt such solution in order to include transcendental operations in calculation, supported differently by different computing architectures. Such operations would not be tested 
using typical production kernels, as, for instance, the Kaiser-Bessel function \citep{jackson_91}, usually tabulated and interpolated. On the other hand, analytical approximations like \cite{Barnett2019APN}, are computationally equivalent to the Gaussian. In any case, different kernels are not expected to  introduce substantially different computational overheads, as also stated by \cite{offringa-wsclean-2014}. 
The last sub-step is represented by the management of the data exchange among different computing units. 

The third part of the algorithm performs the Fast Fourier Transform (FFT) of the gridded data, producing the real space image. This is performed using the FFTW library.

The fourth step is the application of the phase shift and the reduction of the $w$-planes into the final image. 

In the fifth and final step the resulting image is written in a file on the disk. 

\section{The HPC implementation}
\label{hpc-implementation}

\begin{figure}
\includegraphics[width=0.48\textwidth]{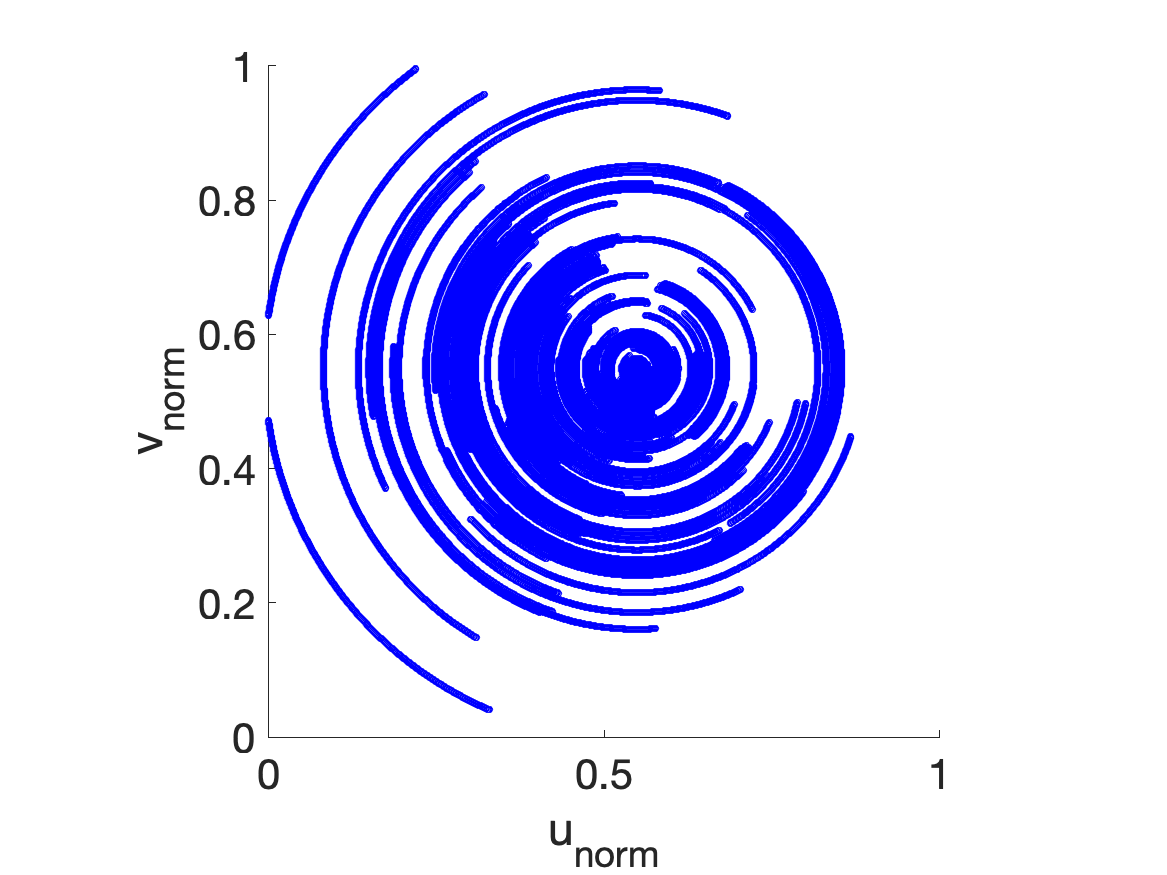}
\caption{Example of measurements taken by a radio-interferometric telescope. A 8 hours observation is shown. Both the angular coverage and the radial distribution can be significantly uneven. Both $u$ and $v$ coordinates are arbitrarily normalised.}
\label{fig:coords}
\end{figure}

The HPC implementation of our w-stacking gridder consists in: {\it i)} distributing the data among different computing units, {\it ii)} distributing the workload among different computing units, {\it iii)} speeding-up the work exploiting accelerators. Throughout the paper, we refer to {\it computing or processing unit} as the computing entity addressing some parts of the work. In the case of parallel work based on MPI, a computing unit is a single core (mapping to an {\it MPI task}). In the case of multithreaded OpenMP implementation, it is a multicore CPU. In the case of accelerated computing, it is a GPU.

\subsection{Data distribution}
\label{data_dist}

In order to achieve an effective memory utilisation, all big datasets, in our case those related to measured visibilities and those defined on the mesh, must be distributed. The concurrent presence of unstructured point like data (the observed visibilities), together with regular mesh based data (the convolved visibilities) requires some care in order to obtain an efficient data distribution. 
\smallskip

\noindent {\bf Measurement data distribution} -- Point like visibilities are collected as time series during an observation. At each measurement time (snapshot), data coming from all pairs of antennas of the interferometer (baselines) are collected. Snapshots are recorded one after the other. The resulting visibility distribution has a spatial number density that varies dramatically depending on the arrangement of the antennas and on the observation time. For instance, areas around $(u,v)=(0,0)$  are denser than peripheral areas for arrays with a core-dense layout. Also radial number density is not uniform, changing with the orientation. Figure \ref{fig:coords} shows examples of such kind of uneven distributions.

\begin{figure}
\includegraphics[width=0.48\textwidth]{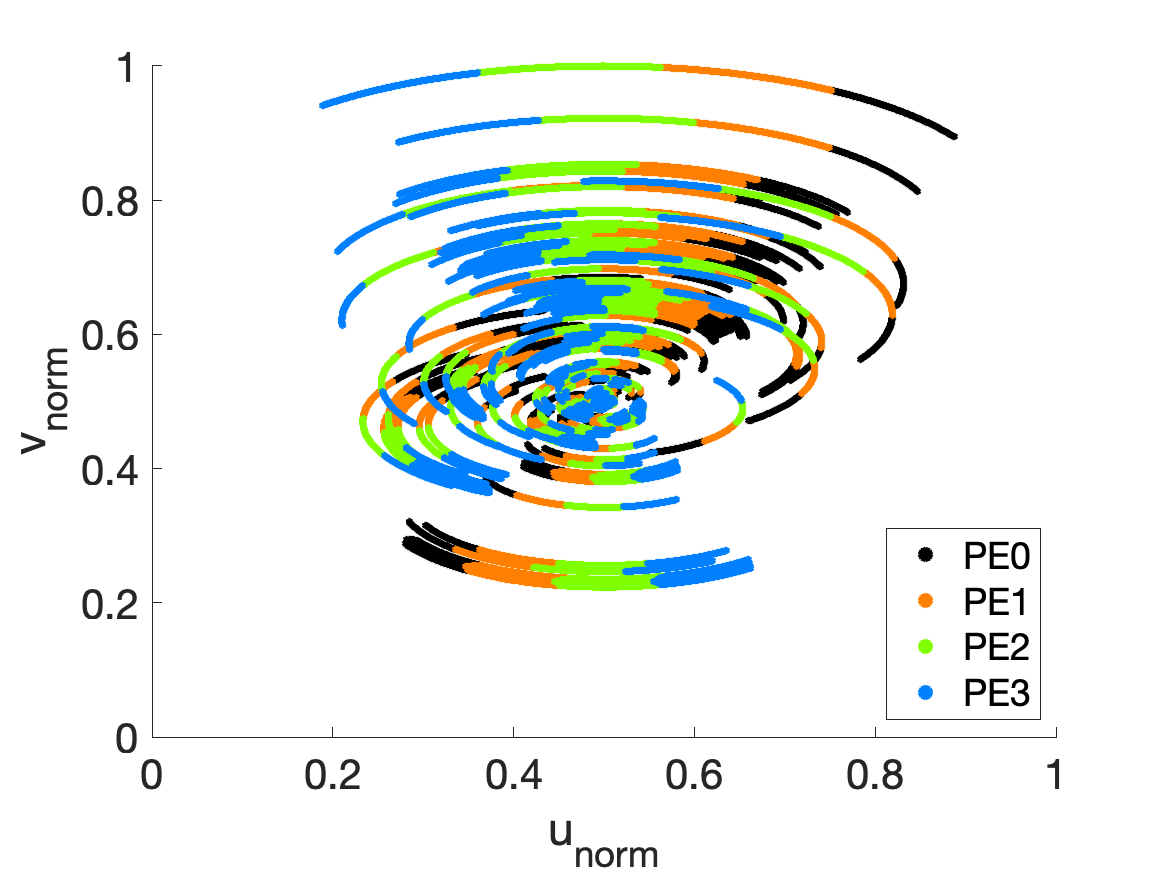}
\caption{2D Distribution of the observational data among 4 processing units (indicated as PE\#). Subsequent time slices are assigned to each processor. Each processor owns data distributed on the whole $(u,v)$ plane.}
\label{fig:coords_pe}
\end{figure}

In our approach, visibility data is distributed among processing units in time slices. If $t_{obs}$ is the total observation time and $N_{pu}$ is the number of processing units, we divide the data into $N_{pu}$ time slices each comprising data in a time interval equal to $t_{pu} = t_{obs}$/$N_{pu}$. Processing unit 0 owns the data from time 0 to $t_{pu}$, processing unit 1 from $t_{pu}$ to 2$t_{pu}$, and so on. An example of data distribution among 4 processing units is presented in Figure \ref{fig:coords_pe}. 

This data distribution has several advantages. First and foremost, data distribution is balanced, because each computing unit owns the same number of visibilities, which are spread on the whole $(u,v,w)$ volume. Second, data can be distributed already when it is read from the disk, without subsequent communication. This, of course, requires also a parallel file system (which we assume to be available in an HPC facility) to be efficient. Third, data lying on the same area of the Cartesian grid where the visibilities are convolved are distributed among different computing units, reducing the computational pressure and potential race conditions (i.e. concurrent write access of multiple threads on the same grid point of the Cartesian mesh) on that specific area. Main drawbacks are represented by the non locality of the data, that are spread all over the spatial domain, hence the computational mesh, and possible unbalances for short observations. These issues could be alleviated, e.g., by sorting the data according to appropriate criteria. However this would introduce additional computational and communication overheads, counterbalancing the possible benefits \citep[see e.g. ][]{2016CoPhC.207...69H}. Hence we have chosen to adopt the linked-list based approach described in Section 3.2.

\begin{figure}
\centering
\includegraphics[width=0.49\textwidth]{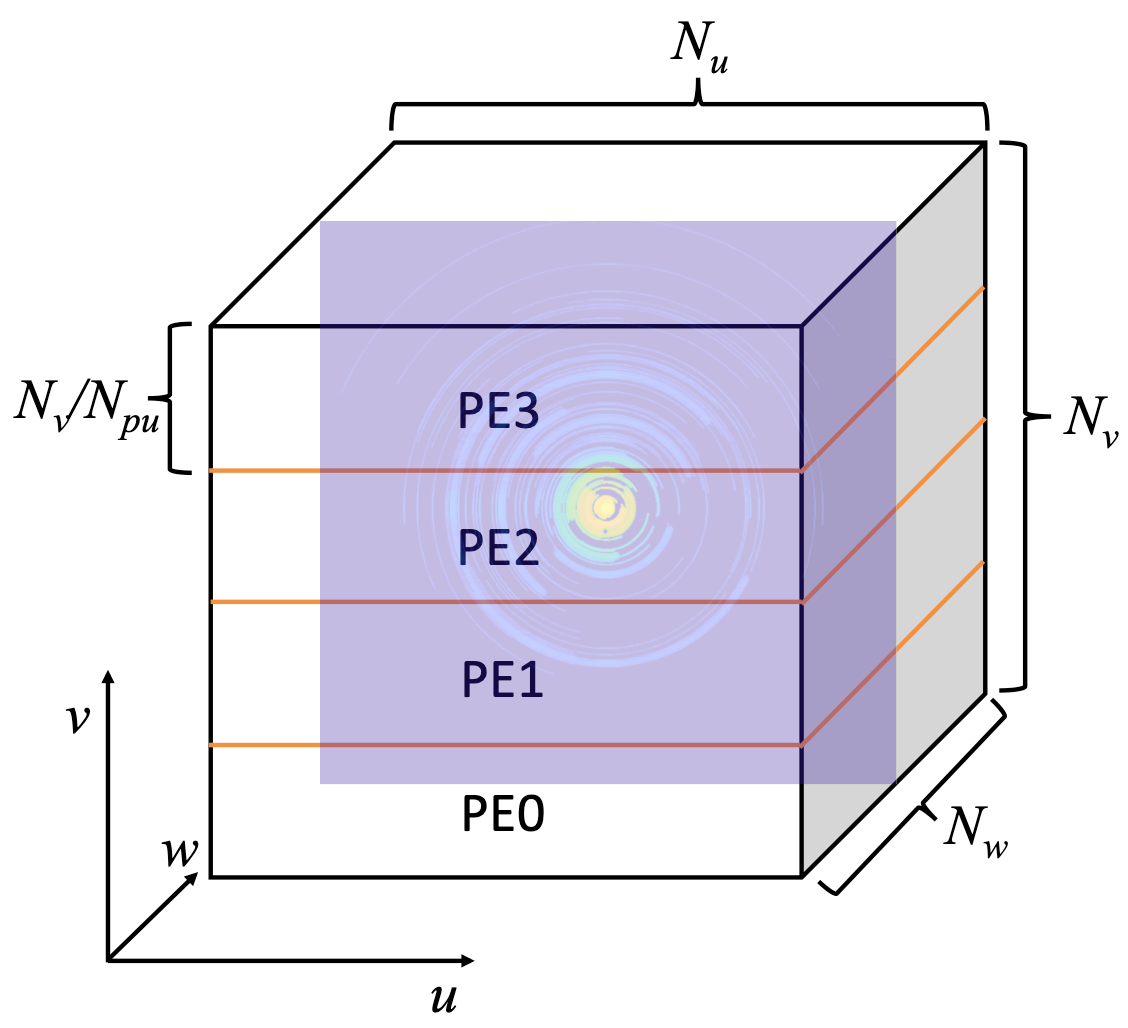}
\caption{Cartesian computational mesh data distribution. Each computing unit (indicated as PE\#) owns a rectangular {\it resident slab} of size $N_u\times N_w\times (N_v/N_{pu})$ cells. In this example data is split among $N_{pu} = 4$ computing units.}
\label{fig:cube}
\end{figure}

\smallskip
\noindent {\bf Mesh data distribution} -- The data defined on the Cartesian computational grid has been partitioned among the computing elements adopting a rectangular slab-like decomposition (Figure \ref{fig:cube}), assigning to each task a rectangular region of $N_{mesh} = N_u \times N_w \times (N_v/N_{pu})$ cells, where $N_u$ and $N_w$ are the mesh size in the u and in the $w$ directions respectively. $N_v$ is the mesh size in the v direction, that is split in $N_{pu}$ parts, equal to the number of computing tasks. 

Such domain decomposition allows storing on each memory only a fraction of the whole mesh, avoiding replicas. Hence, each different computing element is in charge of storing and managing a specific part of the mesh (hereafter the {\it resident slab}). Furthermore, slab decomposition represents a natural layout for the usage of the parallel FFTW library, to transform mesh based visibility data to images. The main drawbacks of this approach are that {\it i})  the maximum number of computing units that can be used is equal to the number of cells in the v-direction (along which the mesh is decomposed): $N_{pu}\le N_v$. Since however we deal with rather big $N_v > 1000$ meshes this is not expected to be a substantial limiting factor;
{\it ii}) a suitable communication scheme has to be implemented in order to recompose the images.

\subsection{Parallel Processing}
\label{mpi}

Data distribution implies additional management of the work performed by multiple computing units and some communication to communicate data among processors. Furthermore, synchronisation has to be properly enforced in order to ensure that communication and work are properly completed at specific steps of code execution, when this is necessary (e.g. before writing results on disk). Communication and synchronisation limit the effectiveness of the parallel execution of an algorithm, i.e. its scalability, which measures the ability to handle more work as the size of the computer or of the data grows. Therefore, they both have to be minimised in order to exploit larger and larger computing resources, supporting bigger data and reducing the time to solution. This is accomplished by exploiting the Message Passing Interface \citep[MPI,][]{Gropp:1994:UMP}, the de-facto standard for distributed parallel computing.

\begin{figure*}
\centering
\includegraphics[width=0.8\textwidth]{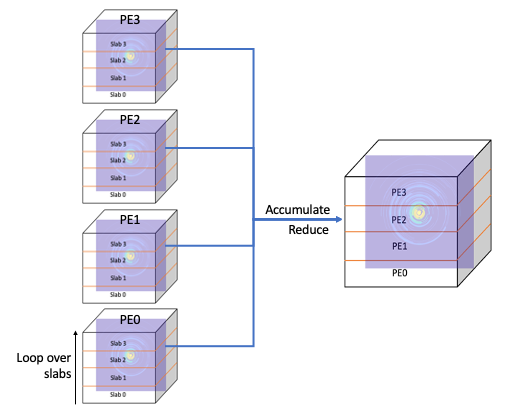}
\caption{The different computing units (four units in this example, indicated as PE\#) perform the gridding operation on their respective fraction of visibilities slab by slab. The process starts from slab 0: all processing units calculates their local contribution to slab 0 exploiting the linked list to accelerate visibility data access in memory. At the end, the contribution of each computing unit to slab 0 is {\it accumulated/reduced} (i.e. summed) on the resident slab, which is stored in PE0 memory. Then the process is repeated for slab 1 (accumulating on the resident slab on PE1), for slab 2 (as in this example, accumulating on the resident slab on PE2) and, finally, for slab 3 (accumulating on the resident slab on PE3).}
\label{fig:accumulate}
\end{figure*}

\smallskip
\noindent {\bf Data I/O} -- Data read from disk is implemented by exploiting standard POSIX I/O on the top of a parallel file system. Each computing unit opens concurrently the input binary files and reads the assigned part of data (fseek, fread functions are used). Performance of the I/O subsystem has not been targeted by our work, therefore no specific performance analysis and optimisation have been carried out. The same holds for the writing operation of the final image. MPI I/O could be adopted to improve the performance of these operations if necessary. 

\smallskip
\noindent {\bf Visibility mesh compositing} -- 
The Cartesian computational mesh is split into slabs, as described in Section \ref{data_dist}, that are processed according to the following procedure.

Each processing unit concurrently calculates the local contribution of its own visibilities lying on a given slab. This local contribution is stored by each processing unit in an auxiliary buffer of the same size of the resident slab. As soon as visibilities of the given slab are processed, the local contribution stored onto the auxiliary slab is summed to the relevant resident slab on the processing unit to which it belongs to (resident slab i belongs to PEi)

The parallel summation is implemented adopting two different strategies: {\it i)} using a blocking {\it MPI\_Reduce} function from all the MPI tasks to that owning the resident slab, that implies that all communication has to be completed before the next iteration starts; {\it ii)} using a non-blocking one-side {\it MPI\_Accumulate} operation, designed to perform concurrent reduce operations on the memory of a target processor. Synchronisation is imposed only at the end of the loop over slabs, calling a collective {\it MPI\_Win\_fence} function.

This is performed for all the slabs in a iterative procedure, starting from the slab 0 up to the slab $N_{pu-1}$. Once the last slab has been processed, each processing unit owns its up-to-date resident slab (i.e. with the contribution from the visibilites of the entire domain), ready for the following step (i.e. FFT). Figure \ref{fig:accumulate} sketches four compute units concurrently performing the gridding operation on their fraction of visibilities mapped on slab 2. Once the gridding has been performed, collective communications are used to accumulate or reduce the contribution of each processing unit on the resident slab on process 2.

It is worth highlighting that visibilities lying on the same slab (and belonging to the same processing unit) are scattered in memory, since there is no correlation between the memory location of a measurement and its physical spatial position, i.e. measurements that are close in memory can be spatially distant (and vice-versa). This can lead to an highly inefficient memory usage. In order to speed-up the memory access, prior to the loop over slabs, each processing unit creates $N_{pu}$ linked lists. Each linked list files the local visibilities lying on a given slab. At each slab iteration, the corresponding list is traversed fetching only the data of interest. Fetched data are stored into temporary arrays that, being contiguous in memory, can be efficiently processed by the gridding kernel.

\smallskip
\noindent {\bf FFT and Phase Correction} -- The calculated resident slabs are processed through MPI parallel FFT, using the FFTW library. Each $w$ plane is computed separately. Finally, the w-term phase-shift correction is applied, summing all the different $w$ planes to generate the final 2D image. The phase correction is local and does not require any communication among MPI tasks.

\subsection{Scaling Laws}
\label{scalinglaws}

The scaling laws governing the parallel implementation of our code, can be modelled considering the critical model parameters, namely the number of visibilities $N_{vis}$, the number of computing units $N_{pu}$, and the Cartesian mesh size $N_{\rm I} = N_u \times N_v \times N_w$. In particular we focus on the parts of the code that have been subject to a distributed parallel implementation, namely, the gridding kernel, the FFT, the w-correction and the parallel sum (indicated as ``reduction'').
Each of those functions has a computational cost, in this context defined as the amount of time required to complete certain operation, $C_i$, that is a function of some of the model parameters.

The overall computational cost is:
\begin{equation}
\begin{split}
    C(N_{\rm vis}, N_{\rm I}, N_{pu}) = \alpha_{\rm grid} \times C_{\rm grid}(N_{\rm vis}, N_{pu})+ \\ 
    \alpha_{\rm FFT} \times C_{\rm FFT}(N_{\rm I}, N_{pu}) + \\ 
    \alpha_{w} \times C_{w}(N_{\rm I}, N_{pu})+ \\ 
    \alpha_{\rm red} \times C_{\rm red}(N_{\rm I},N_{pu}),
\end{split}
\label{eq:cost}
\end{equation}
where $\alpha_{\rm i}$ 
are weights that do not depend on the model parameters. 

The computational 
cost of the FFT is \citep{cooley65}:
\begin{equation}
 C_{\rm FFT}(N_{\rm I}, N_w, N_{pu}) \propto \frac{N_{\rm I}}{N_{pu}} \log \left( \frac{N_{\rm I}}{N_{pu}N_w}\right).
\end{equation}

The w-correction cost is:
\begin{equation}
 C_{w}(N_{\rm I}, N_{pu}) \propto \frac{N_{\rm I}}{N_{pu}}.
\end{equation}

Similarly, the cost of the gridding kernel is given by:
\begin{equation}
 C_{\rm grid}(N_{\rm vis}, N_{pu}) \propto \frac{N_{\rm vis}}{N_{pu}}.
\end{equation}
It is worth noting that the parameter $\alpha_{\rm grid}$ depends on the square of the size of the smoothing kernel support (in our work 7 cells), that therefore has a relevant impact on the computational cost of the algorithm. 

Finally the cost of the reduction operation is
\begin{equation}
 C_{red}(N_{\rm I}, N_{pu}) \propto N_{\rm I}\times C_{\rm MPI}(N_{pu}).
\end{equation}
The linear dependency on the mesh size, is due to the fact that in our implementation the total size of the communicated data in the reduction step does not depend on the number of processing units. The factor $C_{\rm MPI}(N_{pu})$ depends on the summation algorithmic strategy (reduce vs. accumulate), on the bandwidth, on the latency and the load of the network and on the MPI implementation of the sum operation. Hence, its precise calculation depends on the local setup and system usage conditions. It is anyway bigger than 1, progressively reducing the efficiency of the parallel sum with increasing the number of MPI tasks. 

Overall, the computational cost depends linearly on the amount of observational data resident on each processing unit and linearly on the slab size. For a given problem size the corresponding contributions to the computational cost decrease with increasing the number of processing unit. On the contrary, the reduce term grows with $N_{pu}$, hence its impact on the computational cost becomes progressively higher, possibly resulting dominant when a high number of MPI tasks is active and/or for small meshes and/or datasets.  

\subsection{Multi/Many-threads parallelism}
\label{mthreads}

The time to solution for each single MPI task can be considerably reduced by accelerating the code exploiting multi-many cores solutions. These solutions are represented either by multi-core CPUs (multi threads parallelism) or by {\it accelerators} like Graphic Processing Units (GPU, many threads parallelism). Accelerators can be adopted as general purpose co-processors coupled to CPUs to speed-up parts of the computation ({\it kernels}) that can be efficiently performed by independent tasks. 

The gridding and the phase correction algorithms have been targeted for multi/many threads parallelism. Convolution and {\it w}-correction of each measurement can in fact be performed independently from all the others with a high arithmetic intensity, representing an ideal case for multiple threads parallel execution.  
The main potential bottleneck is represented by the accumulation of the contributions of each measurement on the computational mesh, that leads to frequent race conditions.  

\subsubsection{Multi threaded CPU implementation}
\label{openmp}

Multi-threading for multi-core CPUs has been implemented by using the OpenMP\footnote{https://www.openmp.org} application programming interface (API). OpenMP represents a consolidated and stable solution to exploit shared memory devices, ensuring good scalability up to many cores and high portability, being supported by all the major hardware and software providers and implemented in all the most common compilers of, in particular, Fortran, C and C++ programming languages. It is based on a set of directives which instruct the code on how to split the work among the available threads, with minimum impact on the source code, although specific customisation has to be done in order to ensure good performance and scalability.

The OpenMP directives are ignored (interpreted as comments) by the compiler if OpenMP is not explicitly enabled at compile time (for instance, for the GCC compiler this is achieved by means of the \texttt{-fopenmp} compilation flag).

In our code, OpenMP is used to parallelize the gridding and phase correction loops, the former over the measurements, the latter over the grid cells. In both cases, loop iterations are independent from one another and they can be trivially distributed among OpenMP threads. When it comes to accumulation of the contributions from different threads, one has to take care about race conditions (i.e. several threads updating the same memory location simultaneously). We then use the OpenMP built-in {\it atomic} clause. 

\subsubsection{GPU implementation}
\label{gpu}

The approach adopted for the GPU is similar to that described for multi threading. 

For the gridding algorithm, each measurement is assigned to a different thread, that performs the convolution of all frequencies and polarisations. Required data is contiguous in memory, hence effectively coalesced, leading to an efficient memory usage. Having million of measurements a high occupancy of the GPU can be  achieved. 

In the case of the phase correction kernel, each iteration of the nested loops in the three coordinate directions is assigned to a different thread. 

The code implementation adopts two different approaches, the first based on CUDA\footnote{https://developer.nvidia.com/cuda-zone}, the second on the offload constructs of OpenMP. The former is a programming model developed by NVIDIA for general computing on GPUs. It is currently the most effective and comprehensive approach to program graphic accelerators. However its usage is restricted to NVIDIA GPUs limiting the software portability. 
The latter consists in a set of directives introduced starting from OpenMP version 4.5, allowing developers to offload data and execution to target accelerators such as GPUs. The usage of a directive based approach hides some of the complexity intrinsic to a procedural approach like CUDA. The compiler is in charge of translating the kernels decorated with directives into constructs addressing the accelerator. This, at the expense of some control, which limits the maximum achievable performance. Furthermore, although in principle OpenMP is portable across different platforms and architectures, it is still supported by a limited number of compilers. 
GCC, version $>$ 9, currently supports the full standard and it provides accelerated codes for NVIDIA and AMD GPUs.
NVIDIA recently released an  HPC standard development kit, that includes C and Fortran compilers and a full support for a set of programming models including OpenMP, OpenACC and CUDA.

\smallskip
{\noindent\bf CUDA implementation --} In order to maintain a single source code and to preserve its portability, the sections of the code specific to the GPU implementation are activated at compile time exploiting the \texttt{\_\_CUDACC\_\_} macro, that is automatically defined by the CUDA NVCC compiler. 

If such macro is defined, the two functions targeting the GPU:
\begin{itemize}
\item include the management of the data, copying to/from the GPU those (and only those, in order to minimise data transfers) arrays necessary for the computation;
\item switch the loop over visibilities and those over the three coordinate directions to GPU kernel calls, assigning each iteration to a different thread;
\item manage the race conditions updating the shared arrays by means of the \texttt{atomicAdd} CUDA operation, which has hardware support to optimise the concurrent write-access to memory locations.   

If not compiled with NVCC, the compiler ignores the CUDA sections and the code runs on the CPU. 

\end{itemize}

\smallskip
{\noindent\bf OpenMP implementation --} OpenMP  directive--based programming model allows the users to insert non-executable \texttt{pragma} constructs that guide the compiler to handle the low-level complexities of the system. 

Currently OpenMP  offers a set of directives for offloading the execution of code regions onto accelerator devices. The directives can define target regions that are mapped to the target device, or define data movement between the host memory and the target device memory. The main directives we are implementing are \texttt{target data} and \texttt{target}, which create the data environment and offload the execution of a code region on an accelerator device, respectively. Both directives are paired with a map clause to manage the data associated with it; the map clause specifies the direction (to, from, or tofrom) of the data movement between the host memory and the target device memory. 

In case of multiple GPUs available on the same node, a flag is used to map each GPU to a different MPI task, in this way we optimise the use of the available resources. The \texttt{\_\_ACCOMP\_\_} macro, is used to activate at compile time some OpenMP specific functions (e.g. inspecting the GPUs hardware and number of available GPUs).

Race conditions are addressed using the same {\it atomic} directive used for the CPU version.

The compiler used for all the OpenMP tests is the NVIDIA Standard development kit NVCC compiler, the same framework used for the CUDA tests.

\section{Results}
\label{results}

In order to analyse the performance and the scalability of the HPC implementation of our code, we have performed a number of tests and benchmarks exploiting two different state-of-the-art HPC architectures, available at J\"ulich and CINECA supercomputing centres. 

The Forschungszentrum J\"ulich operates the Juwels Booster Module, with 936 2$\times$AMD EPYC 7402 24C 2.8GHz nodes equipped with 4 NVIDIA A100 GPU, 40 GB HBM2e each, connected by a 4$\times$InfiniBand HDR interconnect. CINECA is the Italian national HPC facility running the Marconi100 system, made of 980 2$\times$16-cores IBM Power9 AC922 CPU nodes equipped with 4 NVIDIA Volta V100 GPUs. Each CPU node has 256 GB of DDR4 memory, the memory of the GPU is a 16 GB HBM2. The CPU and the GPU are interfaced by a PCIe Gen4 interconnect, while the 4 GPUs per node adopts NVLink 2.0. The system interconnect is a Mellanox IB EDR DragonFly++. Both systems deploy a IBM Spectrum Scale parallel filesystem.

Table \ref{tab:system} summarizes the characteristics of the systems and indicates the version of the adopted compilers and MPI libraries.

\begin{table*}
\centering \tabcolsep 2pt
\begin{tabular}{l|l||l|l|l|l}
\hline
\hline
System    & Computing Node & $N_{nodes}$ & Interconnect & Compiler & MPI distribution\\
\hline
\hline
M100   & 2$\times$16-cores IBM Power9 AC922 CPUs & 980 & Mellanox IB EDR DragonFly++  & GCC 8.4.0 & Spectrum MPI 10.4.0\\
 & + 4$\times$NVIDIA Tesla V100 GPUs  &  & & NVCC 10.1  & \\
\hline
JUWELS & 2$\times$ 24 cores AMD EPYC 7402 24C CPUs  & 940 & 4$\times$InfiniBand HDR (Connect-X6) & GCC 10.3.0 & ParaStationMPI 5.4.10\\
 & + 4$\times$NVIDIA Tesla A100 GPUs &  & & NVCC 11.3 & \\
\hline
\hline
\end{tabular}
\caption{Technical characteristics of the two HPC platforms adopted for the tests presented in the paper, the adopted compilers and the corresponding compilation options.}
\label{tab:system}
\end{table*}

Two different input datasets have been used for the tests and the benchmarks, namely the {\it Medium} and the {\it Large} dataset. The two datasets come from a LOFAR HBA Inner Station 8 hours observation, whose main characteristics are presented in Table \ref{table:observation}. The Medium dataset is a single sub-band at 146 MHz with bandwidth of 2 MHz (such narrow bandwidth is peculiar of the specific observation, typical HBA observations having a 48 MHz bandwidth) split into 20 channels. It has been used for benchmarking the single node/GPU implementation of the code and for strong scalability tests. The Large dataset is made of the full observation, consisting of 25 sub-bands of 2 MHz each, spanning a frequency range between 120 and 170 MHz. It is used for weak scalability tests and to produce the ``Large Images" presented in Section \ref{sec:large}. The main features of the two datasets together with the size of the computational mesh adopted in the two cases (increasingly bigger), are presented in Table \ref{table:datasets}.

\begin{table}
\begin{center}
\centering \tabcolsep 2pt
\begin{tabular}{l|l|}
\hline
LOFAR HBA Inner Station \\ 
\hline
RA  & 15:58:18.96 \\
DEC & +27.29.19.2 \\
Observation time & 8 hrs \\
Integration time & 4 sec \\
N. antennas  & 62 \\
Bandwidth & 120 - 170 MHz \\
N. of Sub-bands   & 25\\
N. of channels per Sub-band & 20\\
Sub-band bandwidth & 2 MHz\\
MaxUVdist (wavelenghts) &  58624.99 \\
MaxUVdist (meters) &  119902.23 \\
MinUVdist (wavelenghts) &  13.61 \\
MinUVdist (meters) &  27.83 \\
Field of View & $\sim$12 deg$^2$ \\
Angular resolution & $\sim$6 arcsec \\
\hline
\end{tabular}
\end{center}
\caption{{Main characteristics of the LOFAR HBA observation used for the Medium and Large test datasets. Coordinates of the target are given in the first two rows.}}
\label{table:observation}
\end{table}

\begin{table}
\begin{center}
\centering \tabcolsep 2pt
\begin{tabular}{l|l|l}
\hline
    & Medium & Large \\ 
\hline
N. visibilities (approx)  & 0.54$\times$10$^9$ & 13.6$\times$10$^9$ \\
Input data size (GB)  & 4.4 &  102\\
Mesh size Max ($N_u \times N_v \times N_w$) & 4096$\times$4096$\times$16& 32768$\times$32768$\times$64\\
Mesh size (GB)  & 4.0 & 1024.0 \\
\hline
\end{tabular}
\end{center}
\caption{Characteristics of the input datasets and the computational mesh used in the Medium and Large tests. The mesh size takes into account the total amount of memory required for the real and imaginary part.}
\label{table:datasets}
\end{table}

In all the systems the same source code has been compiled using the GCC and the NVCC compilers and the available MPI library. The only dependency of the code is on the FFTW3 library. The Makefile requires only few adjustments of several variables (e.g. the path to FFTW or to the CUDA libraries) to compile the code. The building procedure requires only a few seconds. 

\subsection{Scalability}
\label{sec:scalability}

\begin{figure*}
\centering
\includegraphics[width=0.9\textwidth]{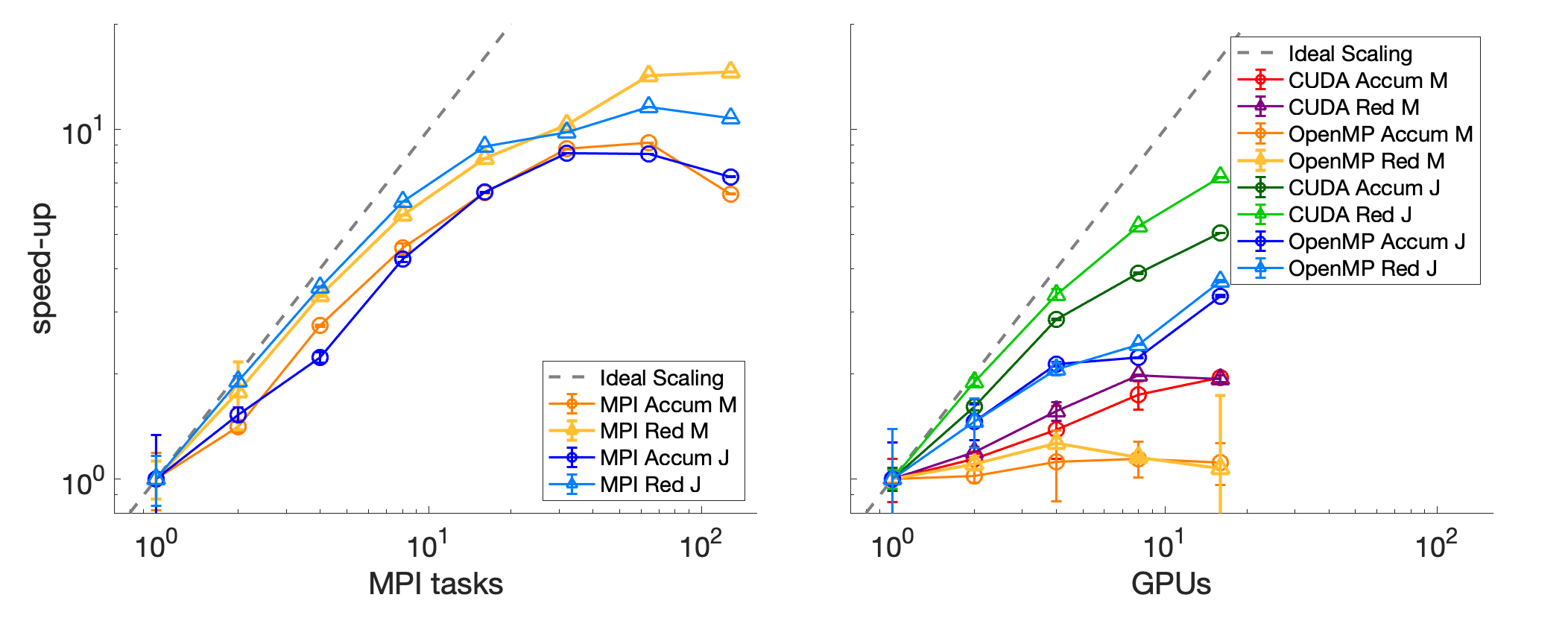}
\caption{Strong scalability of the whole code on the different computing systems: M100 (M) and Juwels (J). Pure MPI runs are shown in the left panel, GPU runs are shown in the right panel. With Accum and Red, the MPI\_Accumulate and MPI\_Reduce versions are indicated respectively. With CUDA and OpenMP the corresponding GPU implementations are indicated. Strong scaling is measured in terms of speed-up (see Equation \ref{eq:speedup}). Ideal linear scaling is shown as a grey, dashed line.}
\label{fig:speedup_total}
\end{figure*}

We first analyse the scalability of our code, broadly defined as the ability to handle more work as the size of the computing system or of the application grows. Strong and weak scalability have been measured.

\subsubsection{Strong Scalability}

A first set of tests focuses on the strong scalability of the code using the Medium dataset (see Table \ref{table:datasets}). Strong scalability measures the performance of the code keeping the data and the mesh size constant and progressively increasing the adopted computational resources (i.e. the number of MPI tasks, the number of OpenMP threads, the number of GPUs). Ideally the code execution time should scale linearly with the number of computing units (i.e. the time should halves if the number of computing units doubles). However, various factors impact such an ideal behaviour, leading to a degradation of the performance with increasing the number of computing units. Therefore, strong scalability indicates the realistic gain the user can expect increasing the number of computing units working cooperatively. 

The strong scalability of the whole code and of its main algorithmic component is measured by the {\it Speed-up} parameter, calculated as:
\begin{equation}
    S = \frac{T_{A,1}}{T_{A,n}}
\label{eq:speedup}
\end{equation}
where $T_{A,n}$ is the wall clock time measured on system $A$ using $N$ computing units and ${T_{A,1}}$ is the corresponding time measured on the same system using a baseline configuration (e.g. a single computing unit). 
OpenMP tests are performed only on a single socket of a computing node (larger configurations being inefficient due to non-local memory accesses), while MPI and GPU tests are performed up to 4 nodes.

\smallskip
{\noindent\bf Strong Scalability of the whole code on the CPU --} The strong scalability of the whole code is presented in Figure \ref{fig:speedup_total}. For both the adopted computing systems, using the CPU cores ({\it pure MPI} runs, left panel) an almost linear scalability is kept up to 32 cores. The speed-up decreases above such number, reflecting the increased impact of communication to the total computing time, due to several issues, discussed in Section \ref{scalinglaws}. In summary: {\it i)} the amount of communicated data does not change with increasing the number of MPI tasks, while the computational load of each task decreases (having to process smaller and smaller domains), {\it ii)} the number of MPI operations increases with the number of tasks, with a bigger traffic overhead; in addition: {\it iii)} above 32 cores communication involves the interconnect, which is anyway slower than local memory access, {\it iv)} unbalances in the code execution grow with the number of tasks, due to the number of data lying on a given slab, which can be different between the different processing units. 

In order to alleviate the impact of unbalances, the MPI\_Accumulate based communication scheme has been introduced. Differently from the MPI\_Reduce based approach, which imposes synchronisation at every iteration sweep over slabs, accumulating overheads due to unbalances, the MPI\_Accumulate approach is non blocking. It requires synchronisation only at the end of the loop over slabs. Since the total number of visibilities is the same on each computing unit, this approach is expected to be unaffected by unbalances local to the slabs.

The scalability curves however show an opposite behaviour, with the MPI\_Accumulate showing slightly worse scalability compared to the MPI\_Reduce. This will be further discussed later in this section, taking into account also the outcomes of the weak scalability tests.\smallskip

\begin{figure}
\centering
\includegraphics[width=0.45\textwidth]{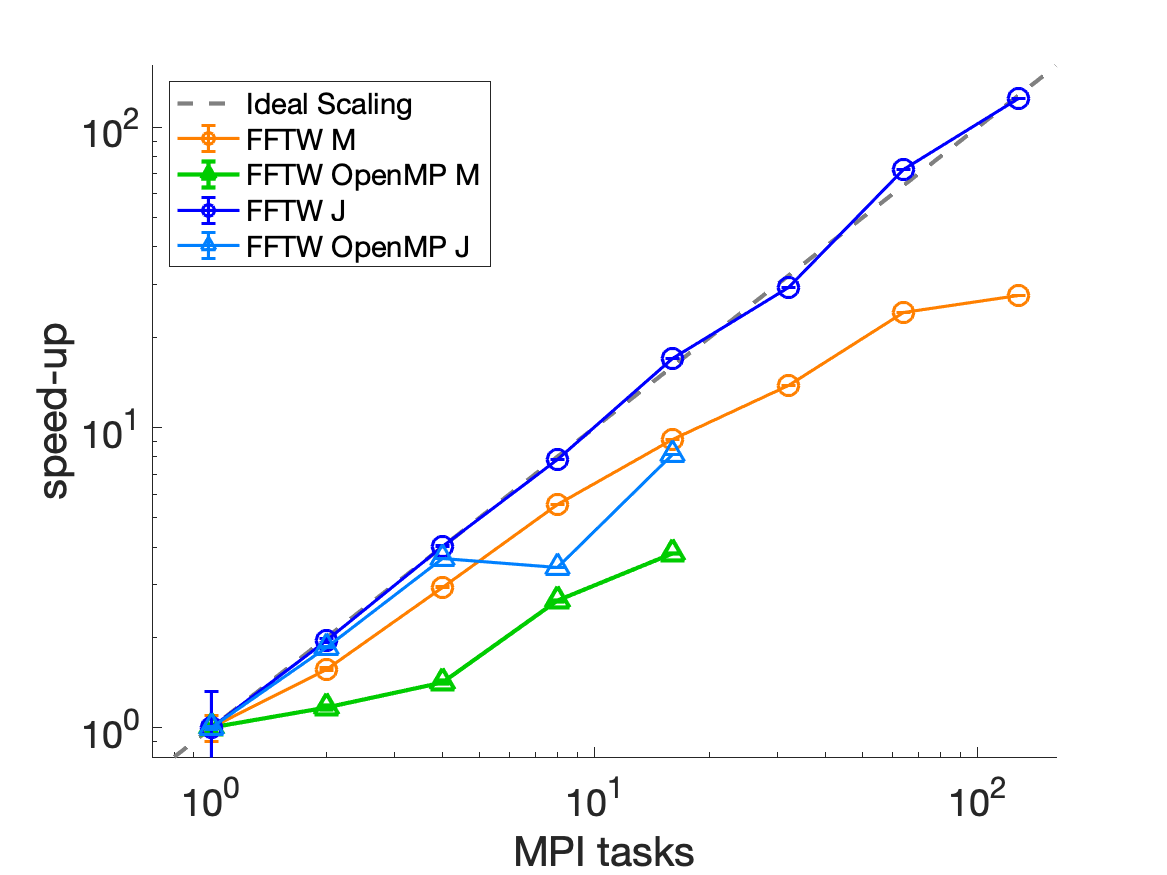}
\caption{Strong scalability of the FFT algorithm on the different computing systems: M100 (M) and Juwels (J). The OpenMP CPU version of the code is indicated as OpenMP. The remaining curves correspond to pure MPI parallelism. Strong scaling is measured in terms of speed-up (see Equation \ref{eq:speedup}). Ideal linear scaling is shown as a grey, dashed line.}
\label{fig:speedup_fftw}
\end{figure}

\smallskip
{\noindent\bf Strong Scalability of the whole code on the GPU --} When GPUs are used (right panel of Figure \ref{fig:speedup_total}), scalability is influenced by the faster execution time of the GPUs compared to the CPUs, and the corresponding bigger impact of communication, that leads to a degradation of the speed-up (see Section \ref{scalinglaws}). Scaling varies depending on the GPU implementation. The CUDA version shows a better scalability than the OpenMP one. The execution of the GPU kernels of the OpenMP version is, in fact, slightly faster than the CUDA one, leading to a higher impact of communication and to a consequent degradation of the  scaling. 
The usage of the FFTW library has a non-negligible influence on the scalability of the GPU version. The fraction of the total computing time spent on the FFTW library is, in fact, relatively higher for the GPU version compared to the CPU one, since gridding and w-correction are performed much faster on the accelerator. 
On Juwels the pure MPI scalability of the FFTW library is linear, preserving a good scaling also for the whole application (see Figure~\ref{fig:speedup_fftw}). The speed-up is instead sub-linear on M100. In the two tests indicated as {\it OpenMP}, FFTW has been recompiled using the NVCC compiler, which produces a less scalable FFTW library. \smallskip

\begin{figure*}
\centering
\includegraphics[width=0.99\textwidth]{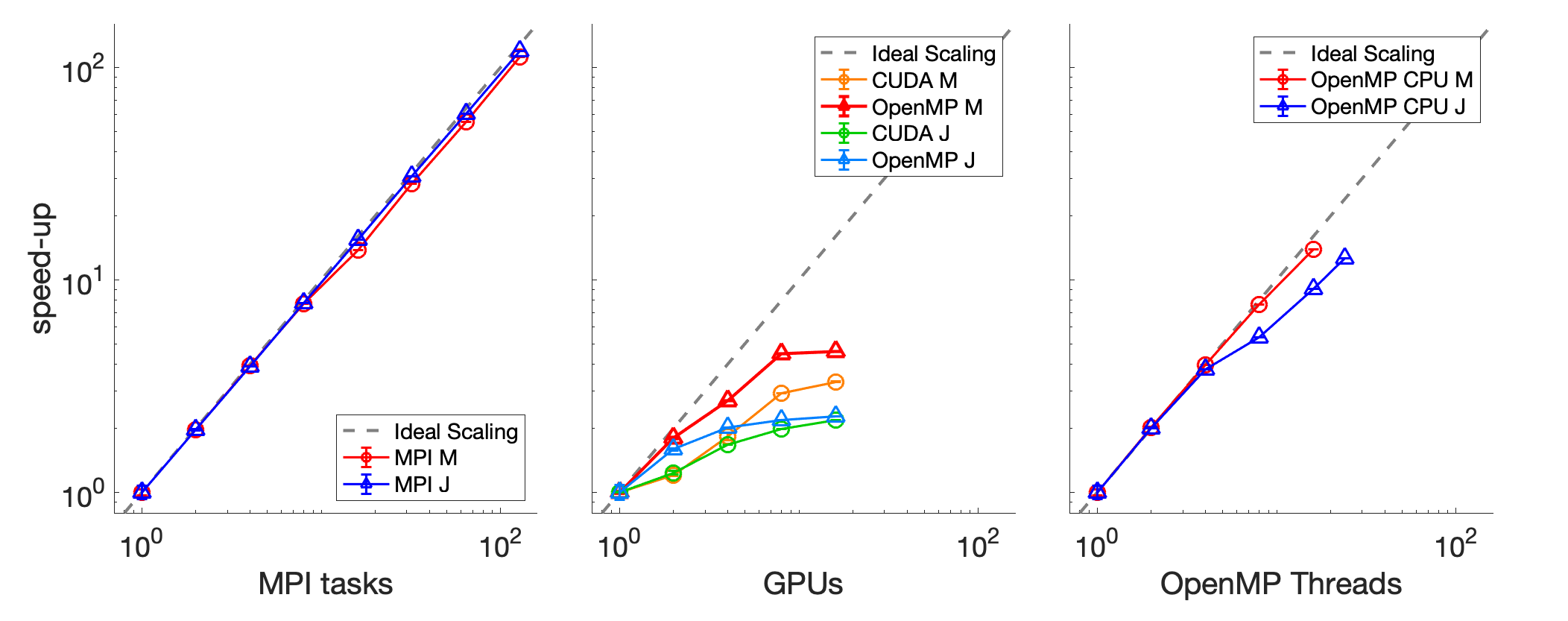}
\caption{Strong scalability of the gridding algorithm on the different computing systems: M100 (M) and Juwels (J). Pure MPI runs are shown in the left panel, GPU runs are shown in the centre panel (with CUDA and OpenMP the corresponding GPU implementations are indicated) and pure CPU OpenMP runs (limited to a single socket, 16 cores for M100, 18 cores for Juwels) are shown on the right panel. Strong scaling is measured in terms of speed-up (see Equation \ref{eq:speedup}). Ideal linear scaling is shown as a grey, dashed line.}
\label{fig:speedup_gridding}
\end{figure*}

\begin{figure*}
\centering
\includegraphics[width=0.99\textwidth]{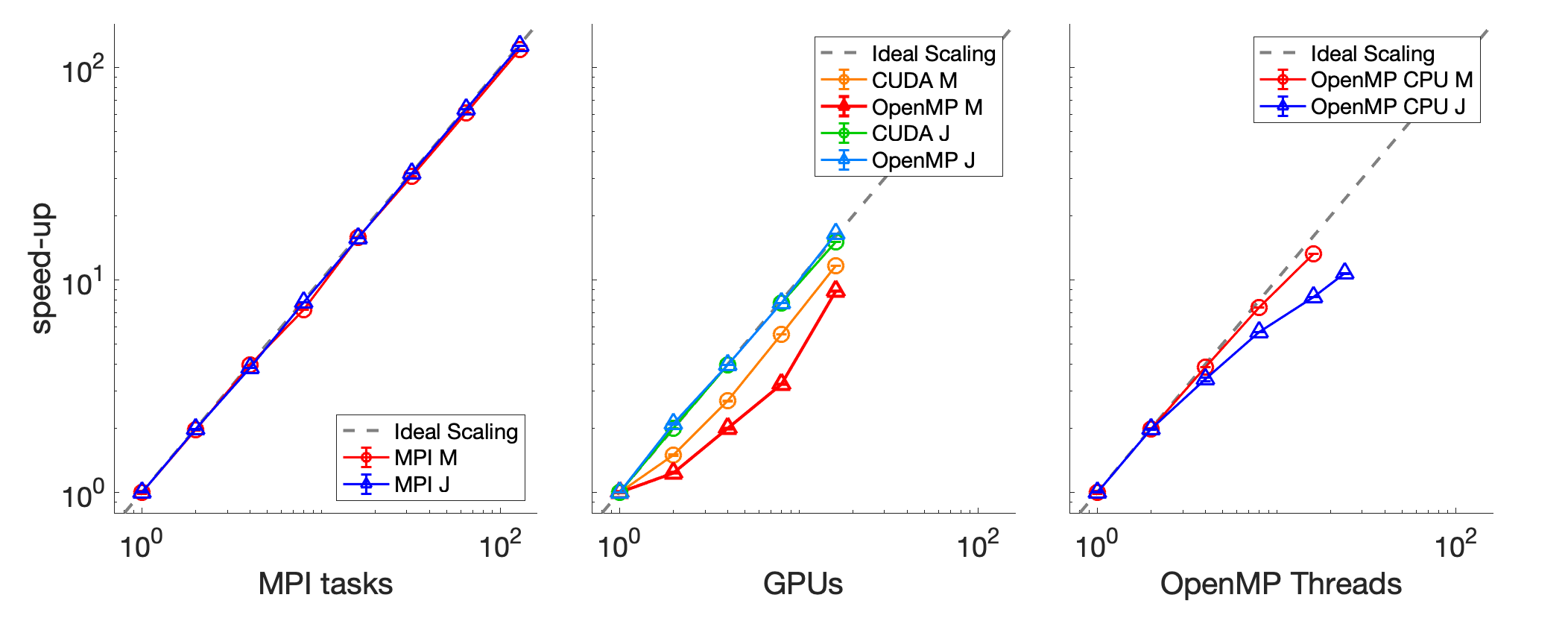}
\caption{Strong scalability of the phase correction algorithm on the different computing systems: M100 (M) and Juwels (J). Pure MPI runs are shown in the left panel, GPU runs are shown in the centre panel (with CUDA and OpenMP the corresponding GPU implementations are indicated) and pure CPU OpenMP runs (limited to a single socket, 16 cores for M100, 18 cores for Juwels) are shown on the right panel. Strong scaling is measured in terms of speed-up (see Equation \ref{eq:speedup}). Ideal linear scaling is shown as a grey, dashed line.}
\label{fig:speedup_phase}
\end{figure*}

{\noindent Strong Scalability of the Gridding and $w$-stacking kernels} -- Figure \ref{fig:speedup_gridding} and \ref{fig:speedup_phase} present the speed-up of those parts of the code that do not imply any communication, namely the gridding and the phase correction kernels. For these kernels the work is evenly distributed among the computing units, as discussed in Section \ref{scalinglaws}. The pure MPI multi-core tests (left panel in both figures), show a perfectly linear scaling in all systems. The central panels presents the speed-up when GPUs are used. Nearly linear scaling is obtained by the phase correction kernel, due to the perfectly data parallel algorithm. For the gridding algorithm, the GPU implementations do not approach a linear speed-up due to the progressive loss of efficiency of the GPU as the size of the amount of data to process decreases. This happens distributing the data among multiple accelerators. 
When 8 and 16 GPUs are used, the V100 architecture keeps increasing the performance (although sub-linearly), while data is not big enough to lead to an efficient usage of the more powerful A100 GPU.

In the right panel of the two figures, the OpenMP multicore scalability is presented. OpenMP data is presented up to the number of cores on a single socket, ensuring data locality through system memory-process bindings. The loop scheduling policy is static in order to exploit memory locality. For both kernels the resulting speed-up is close to ideal, although slightly worse than in the pure MPI case.

\subsubsection{Weak Scalability.} 

Weak scalability has been studied using the Large dataset. In case of weak scalability, both the number of processing units and the problem size are progressively increased, resulting in a constant memory and work load per processing unit. Weak scaling is particularly meaningful for applications where the memory request cannot be satisfied by a single node, as happens for large observational data and/or large images, addressing high resolution and/or wide fields-of-view.

\begin{figure*}
\centering
\includegraphics[width=0.45\textwidth]{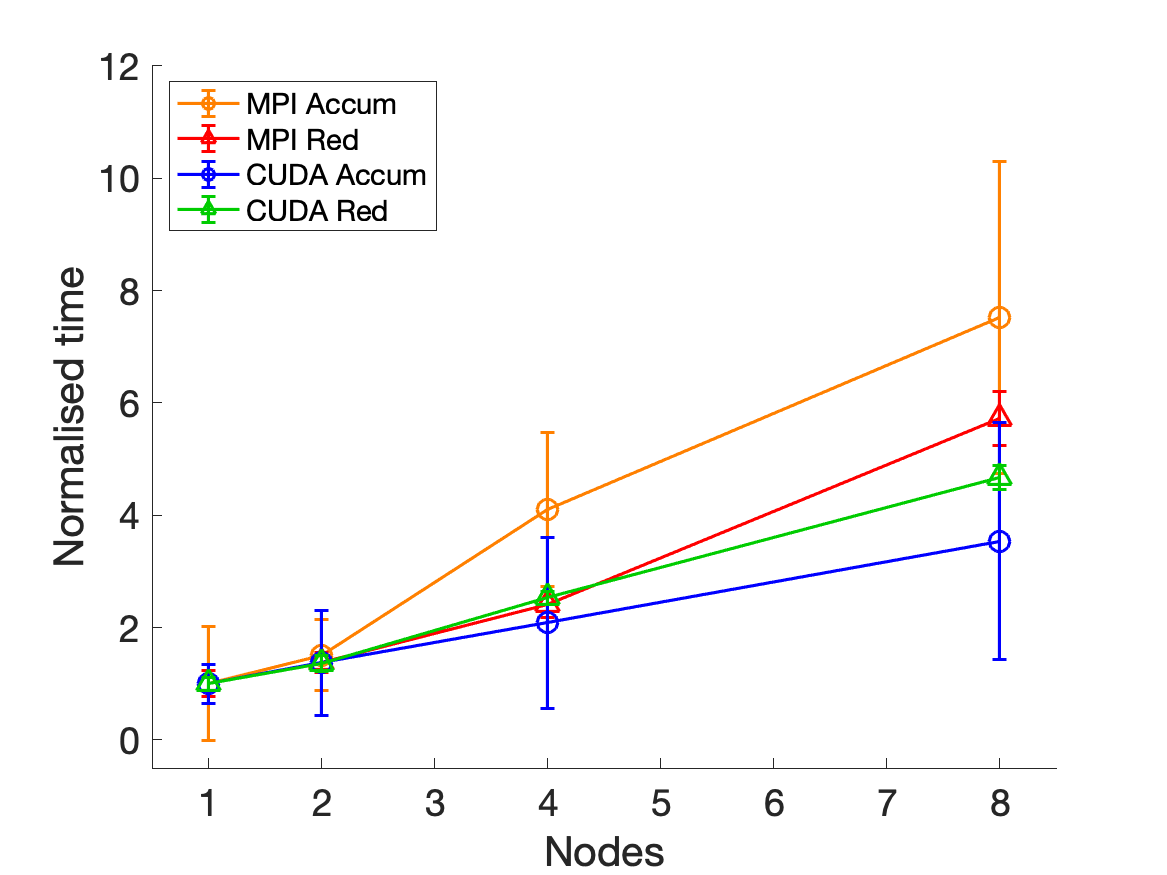}
\includegraphics[width=0.45\textwidth]{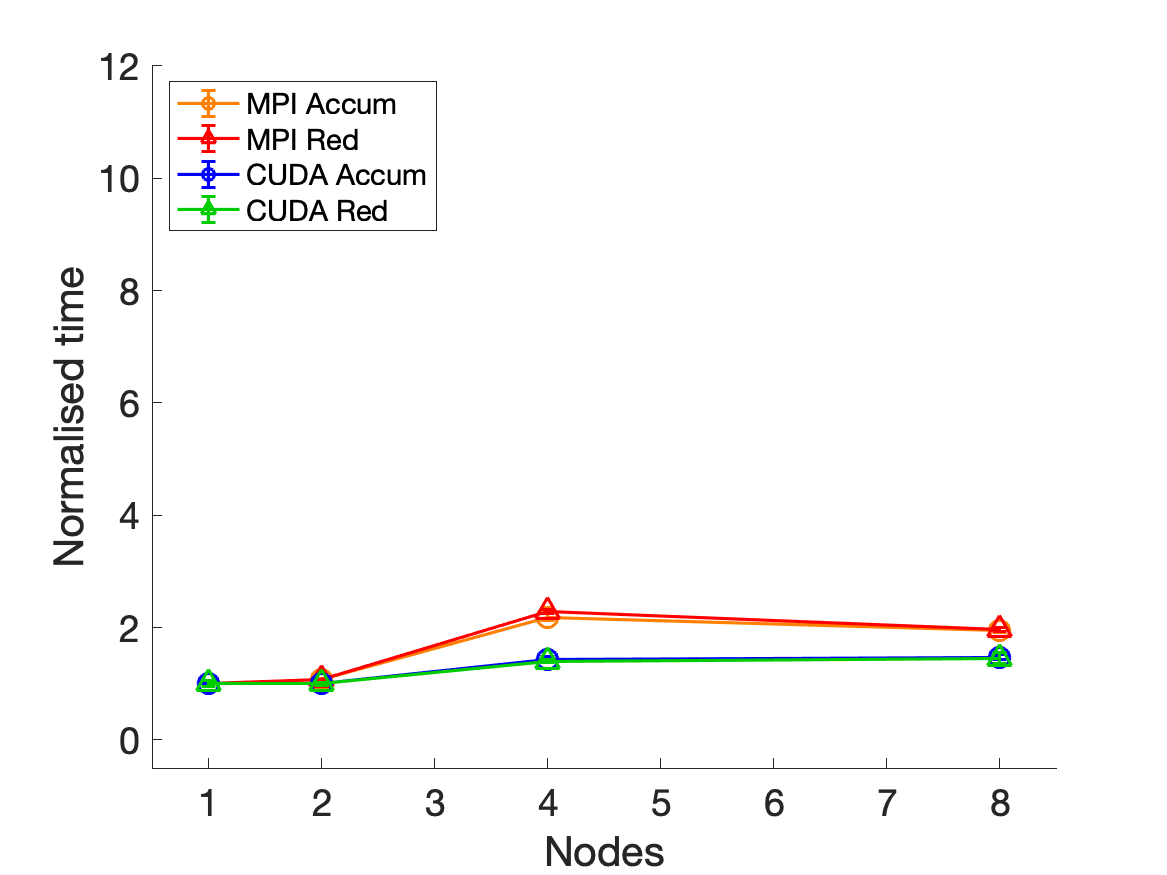}
\caption{ Weak scalability of the whole code (left panel) and of the FFT algorithm (right panel) on the M100 system. With Accum and Red, the MPI\_Accumulate and MPI\_Reduce versions are indicated respectively. With MPI and CUDA, pure MPI runs and GPU runs are indicated.Wall-clock times are normalised to the time to solution on one node.}
\label{fig:weak}
\end{figure*}

Figure \ref{fig:weak} shows the total (left panel) and the FFTW time to solution normalised to the time measured on one node. Timings have been taken on a number of nodes equal to 1, 2, 4, 8, corresponding to a number of cores (and MPI tasks) from 32 to 256 and a number of GPUs (and MPI tasks) from 4 to 32. 

Input data corresponds to the Medium dataset on one node, then they are doubled together with the number of nodes. The largest input dataset, distributed among 8 nodes, is about 36 GB. At the same time, the size of the computational mesh is doubled starting from a 4096$\times$4096$\times$16 setup. The largest mesh is 8192$\times$8192$\times$32 corresponding to a total memory occupancy of 32 GB. 

With an ideal weak scaling, the time to solution should remain constant with the number of nodes. Left panel of Figure \ref{fig:weak} shows that actually the measured total time to solution increases with the problem size and the number of nodes. Such growth is due essentially to MPI related communication overheads, as discussed in Section \ref{scalinglaws}. The FFTW curves, presented in the right panel, are  substantially flat showing ideal scaling. The same holds for gridding and the phase correction kernels (not shown here being perfectly constant). The overhead is therefore due to the gather of the slab data, and depends on the specific adopted communication approach. 

When the MPI\_Reduce collective communication function is used, the pure MPI case shows a larger overhead compared to the GPU case, due to the larger number of MPI tasks (eight times larger), as expected from Equation 4. When the one-side MPI\_Accumulate based solution is used, such difference is even more visible. If few MPI task are used, as in the case of the GPU runs, the accumulate based solution provides the best scaling, thanks to its asynchronous nature. However, for a number of tasks bigger than 64 (2 nodes, for the pure MPI case) the scaling significantly worsen. Despite non-blocking, also the accumulate solution requires synchronisation at the end of the loop over all the slabs. This is implemented through and MPI\_Win\_fence call, deployed by the MPI2 standard. Such a synchronisation proves to introduce a strong overhead above 64 tasks, which appears to be a limiting factor in its usage on large configurations. It does not seem to be due to a specific implementation of the standard, affecting both the SpectrumMPI library available on M100 and the ParaStationMPI library available on Juwels. 

\subsection{Performance and Memory Usage}
\label{sec:performance}

In this Section we discuss the performance of the code on a single computing unit. The wall-clock times to solution for the Medium dataset to generate a 4096$\times$4096$\times$16 pixels image are presented in Table \ref{table:timings}. This is the largest set-up that can run on a single GPU (due to memory constraints). Combined to the scalability information presented above, these figures characterise the computational performance of the discussed algorithm and assist the user in estimating the time needed for the w-stacking gridder within a data processing pipeline fully exploiting diverse and heterogeneous HPC resources. \smallskip

{\noindent\bf Single core performance --} The figures collected on a single core (first row of Table \ref{table:timings}) inform on the time to solution when no parallel processing is adopted. The overall time for the application to run varies from around 87 seconds on Juwels to 130 seconds on M100. The Juwels system has higher performance compared to M100 in the gridding and phase-correction kernels (a factor of more than 2), given to the effective vectorisation provided by the AMD processor architecture. The M100 machine deploys instead a highly efficient FFTW library (3 times faster than the Juwels installation).

\smallskip
{\noindent\bf Single GPU performance --} The CUDA implementation of the gridding and phase correction kernels show the expected speed-up compared to the single core benchmarks. All the timings of the CUDA implementation, presented in the second row of Table \ref{table:timings}, include the data copy time to/from the accelerator. The gridding algorithm is 25 and 30 times faster than a single core on the M100 and Juwels systems respectively. 
The phase correction kernel has an even higher performance boost compared to a single core, with a factor of 75 for the V100 and a factor of 40 for the A100 architectures.

The FFTW library has no GPU implementation. Timings of the FFTW are the same as those of a single core, representing the highest share of the overall computing time. On the Juwels system, the FFTW built with the GCC compiler contributes to more than the 80\% of the the total time. On M100 it contributes to the 55\%. Recompiling the FFTW library with the NVCC compiler and custom options improves its speed of a factor of 1.3 on M100 and of about 2 on Juwels (decreasing however its scalability as discussed in Section \ref{sec:scalability}). The corresponding timings are presented in the third row of Table \ref{table:timings}. where also the timings obtained using the GPU OpenMP implementation of the Gridding and Phase Correction kernels are shown (again, including data copy time to/from the GPU). It is interesting to notice that for our portable implementation, OpenMP results faster than CUDA on both computing systems. This proves that current offload capabilities of OpenMP can reach a good level of performance.

Overall, the presence of the non accelerated FFTW, reduces the  performance gain of the GPU enabled version of code to a factor of 8.5 for M100 and of slightly less than a factor of 2.5 on Juwels compared to a single core, emphasising the importance of porting the whole code on the GPU to fully exploit the power of an accelerated system.

\smallskip
{\noindent\bf Single node performance --} The analysis at the node level, measures the performance achievable by a software capable of exploiting all the computational resources of the available CPU. We present the timings obtained using all the cores on the node (pure MPI, fourth row of Table \ref{table:timings}) together with those obtained using all the GPUs (4 GPUs per node on both M100 and Juwels, presented in the fifth and sixth rows of Table \ref{table:timings} for the CUDA and the OpenMP implementations respectively). For the gridding kernel, the CUDA and the OpenMP implementations are about 1.5 and 3 times faster than the pure MPI one respectively on M100, and 1.5 and 2.2 on Juwels. For the phase correction kernel, we get factors of about 6.5 and 7 on M100 and of about 5 for both CUDA and OpenMP on Juwels. 

Overall, on M100 the GPU implementation of the code is between 1.3 (OpenMP) and  1.4 (CUDA) times faster than the pure MPI one. On Juwels we get factors of about 0.8 and 1. However, the GPU code performance on the node is, once more, strongly influenced by the FFTW that cannot exploit the GPU.\smallskip

The effectiveness of our implementation is expressed in terms of \emph{Giga Grid-Point Additions Per Second} (GGPAS), following the works published by \cite{Romein_2012} and \cite{Merry_2016}. It is worth noting that GGPAS counts the number of updates in registers during the execution of the kernel, not the number of updates onto the grid points in the device global memory (i.e. GGPAS does not depend on the size of the grid). For the Medium test the Giga Grid-Point Additions value is estimated to be equal to 26.65. 
The measured GGPAS follows of course the behaviour of the Gridding kernel, with the best performance obtained by the GPU OpenMP implementation running on the Juwels system, with around 49.60 GGPAS. The GGPAS we get is lower than what achieved in the aforementioned works, where, however, specific optimisations have been accomplished. Although this could be done also for our code, it would be inconsistent with our goal of ensuring portability in terms of both code and performance, hence we have not pushed further the tuning of our implementation. 
\smallskip

\begin{figure*}
\centering
\includegraphics[width=0.60\textwidth]{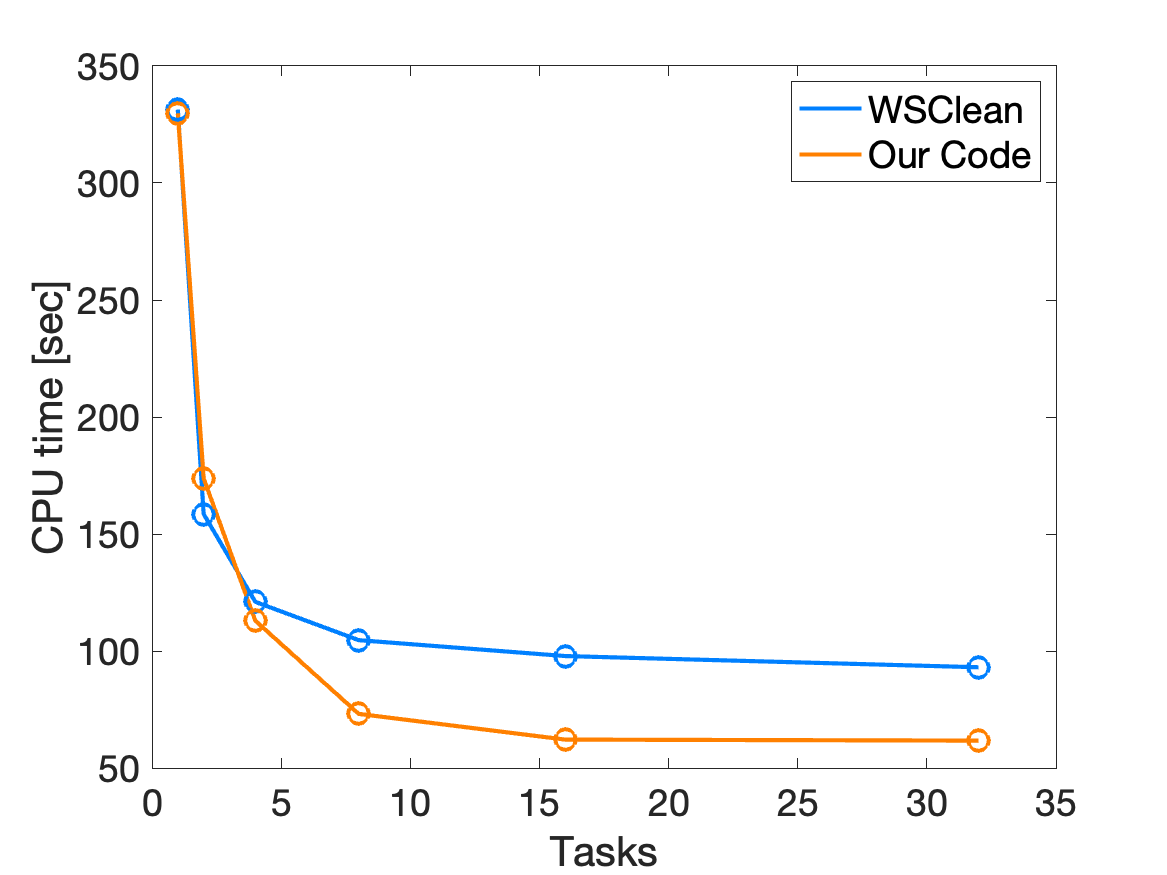}
\caption{Comparison of the CPU time of WSClean and the code presented in this paper (``Our code'') running on the same input dataset with the same set-up, with increasing the number of parallel tasks on a single computing node. For WSClean we have used multithreading while for our code we have adopted MPI parallelism. GPUs have not been used since they are not supported by version 3.0 of WSClean. The same holds for multi-node parallelism.}
\label{fig:wsclean}
\end{figure*}

{\noindent\bf Comparison with WSClean --} 
Figure \ref{fig:wsclean} presents the comparison between the time to solution of our code and one of the state-of-the-art codes for imaging of radioastronomy data, WSClean \citep{offringa-wsclean-2014}. This comparison is meant only as a sanity check of our implementation and not a formal benchmarking, as it is possible that WSClean performance can be improved with careful tuning of its parameters. However, this is outside the scope of this work. WSClean, version 3.0, was run as:

\noindent wsclean -j N -size 4096 4096 -scale 2asec -name test inputdata.ms

\noindent where he option -j sets the number of N threads that the code can use, the -size option sets the image size in pixel, the -scale option sets the resolution in arcsec and the -name option set the prefix for the output files. With this set-up, WSClean performs the same operations accomplished by our code. We have run only on one node and without using the GPUs, in order to have a fair comparison between the two codes. Built-in multithreading parallelism has been used for WSClean, MPI based parallelism has been exploited for our code. We notice that the performance of the two codes is very similar up to 4 tasks. For larger configurations, multithreading scales a bit worse than MPI, leading to a slightly better performance of our code compared to WSClean. Above 16 parallel tasks, both codes do not improve further the performance. For the MPI implementation this is due to the small size of the input dataset, leading to overheads related to parallelism to dominate. Multithreading is instead penalised by a number of architectural issues (e.g. cache coherency, memory contention etc.). 

\smallskip

{\noindent\bf Memory requirements --} The memory footprint of the w-stacking code per computing element (in bytes) can be calculated as:
\begin{equation}
    M \simeq 8\times \frac{1}{N_{\rm pu}}\bigg (3.5\times N_b + (2\times N_{\rm {vis}} + 4\times (N_u \times N_v \times N_w) + L_k \bigg )\ {\rm B},
    \label{eq:memory}
\end{equation}
where $N_b$ is the number of baselines, $N_{\rm {vis}}$ is the total number of visibilities to be mapped to the grid, $N_u \times N_v \times N_w$ is the number of grid cells and $L_k$ is the size of the linked list assigning baselines to the different slabs. The factor 8 reflects the usage of double precision floating point numbers for most of the variables, the factor 3.5 is due to the three phase space coordinates of the baselines plus the single precision weight variable. The factor 2 takes into account the real and imaginary parts of the visibilities and the factor 4 is the product of the factor 2 for the real and imaginary parts of the visibilities calculated on the grid times a further factor of 2 due to the duplication of the slab to keep in memory both the resident slab and the slab used to convolve the visibilities. The size of $L_k$ can be estimated as:
\begin{equation}
    L_k \sim  2\times \alpha \times \frac{N_b}{N_{\rm pu}},
\end{equation}
where the factor of 2 reflects the number of variables characterising a node of the linked list (the id of the baseline and the pointer to the next node of the list), $\alpha$ is a parameter bigger than 1, that accounts for the boundary conditions for each slab (baselines falling in a neighbouring slab but contributing to that being processed). Such parameter depends from the size of the support of the convolution kernel (in our application equal to 7 cells), but it is not expected to be $>2$. For $N_{\rm pu} > 2$, the contribution of $L_k$ to $M$ can be considered negligible, depending from the inverse of the square of the processing units. 

Equation \ref{eq:memory} shows that the size of all the main data structures involved in the computation scales linearly with the number of computing units, memory request halving with doubling the number of parallel tasks. The only meaningful memory overheads, compared to a purely sequential code, are represented by the additional resident slab and the linked list, that, however, has no impact starting from a few computing units. 

\begin{table*}
\begin{center}
\centering 
\begin{tabular}{l|l|l|l|l|l|l|l}
\hline
    & System & Total & Gridding & Phase Corr. & FFTW & Giga-Grid-Point Additions per Second\\ 
    &  & (sec.) & (sec.) & (sec.) & (sec.) & (GGPAS)\\
\hline
CORE  & M100  & 130.80$\pm$0.12 & 81.71$\pm$0.03 & 36.732$\pm$0.014  & 9.27$\pm$0.08       & 0.328$\pm$ 0.002\\
      & Juwels & 87.56$\pm$0.16 & 36.47$\pm$0.04 & 15.839$\pm$ 0.004 & 31.62$\pm$0.11      & 0.733$\pm$0.006\\
\hline      
GPU   & M100  & 15.92$\pm$0.03 & 3.25$\pm$0.01   & 0.493$\pm$0.005  & 9.27$\pm$0.02        & 8.18$\pm$0.02\\
CUDA  & Juwels & 36.79$\pm$0.07  & 1.239$\pm$0.004 & 0.397$\pm$0.001 & 31.55$\pm$0.06      & 20.84$\pm$0.02\\
\hline      
GPU   & M100  & 12.11$\pm$0.01 & 2.64$\pm$0.02 & 0.309$\pm$0.004  & 6.92$\pm$0.03          & 10.09$\pm$0.04\\
OMP   & Juwels & 18.56$\pm$0.39  & 1.08$\pm$0.08 & 0.400$\pm$0.002 & 15.04$\pm$0.01        & 26.15$\pm$0.41\\
\hline      
NODE    & M100  & 12.74$\pm$0.33 & 2.88$\pm$0.01 & 1.193$\pm$0.010   & 0.71$\pm$0.03       & 8.79$\pm$0.02\\
(cores) & Juwels & 8.95$\pm$0.16  & 1.19$\pm$0.01 & 0.499$\pm$0.001 & 1.22$\pm$0.02        & 22.55$\pm$0.04\\
\hline
NODE    & M100  & 8.99$\pm$0.23  & 1.79$\pm$0.06  & 0.181$\pm$0.01  & 3.17$\pm$0.02        & 14.88$\pm$0.18\\
(gpus cuda)  & Juwels & 10.92$\pm$1.36 & 0.738$\pm$0.005 & 0.098$\pm$0.001 & 7.43$\pm$1.38 & 36.11$\pm$0.04\\
\hline
NODE    & M100  & 9.58$\pm$0.07  & 0.98$\pm$0.02  & 0.158$\pm$0.005  & 4.87$\pm$0.03       & 27.06$\pm$0.10\\
(gpus omp)  & Juwels & 9.00$\pm$0.36 & 0.53$\pm$0.03 & 0.100$\pm$0.002 & 4.10$\pm$0.02     & 49.60$\pm$0.29\\
\hline
\end{tabular}
\end{center}
\caption{Timings for processing the Medium dataset generating a 4096$\times$4096$\times$16 pixels image for the different code kernels on different devices (core, GPU, node) and systems (Marconi 100 and Juwels). Errors are calculated as the standard deviation over ten measurements. The last column shows the efficiency in terms of Giga Grid-Point Additions Per Second (GGPAS), i.e. the rate at which the cells of the computational mesh are calculated.}
\label{table:timings}
\end{table*}

\subsection{Large Images}
\label{sec:large}

Our final test is the processing of the Large configuration reported in Table \ref{table:datasets}, with a mesh of 32768$\times$32768$\times$64 and an output image size of 32768$\times$32768 pixels. In order to have the largest possible input dataset, we have combined all the 25 bands of the available observation.

\begin{figure*}
\centering
\includegraphics[width=0.45\textwidth]{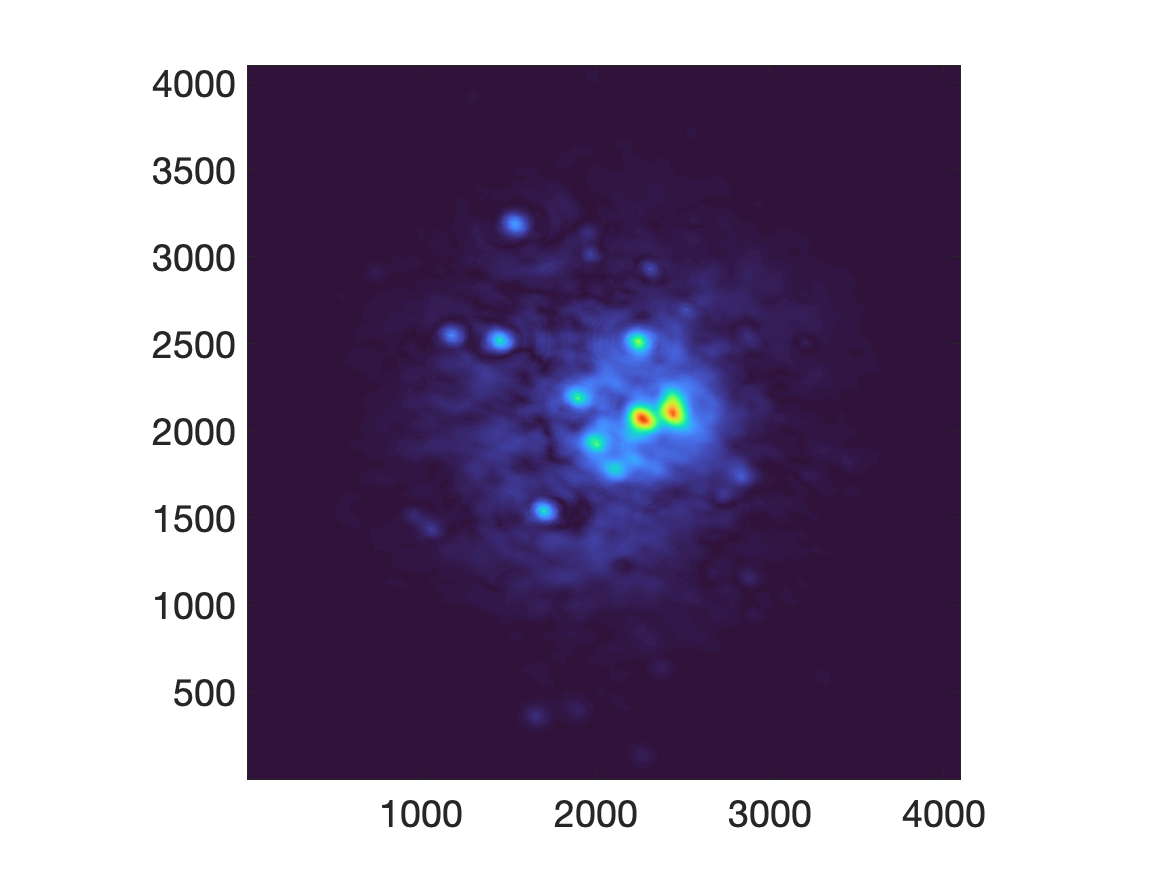}
\includegraphics[width=0.45\textwidth]{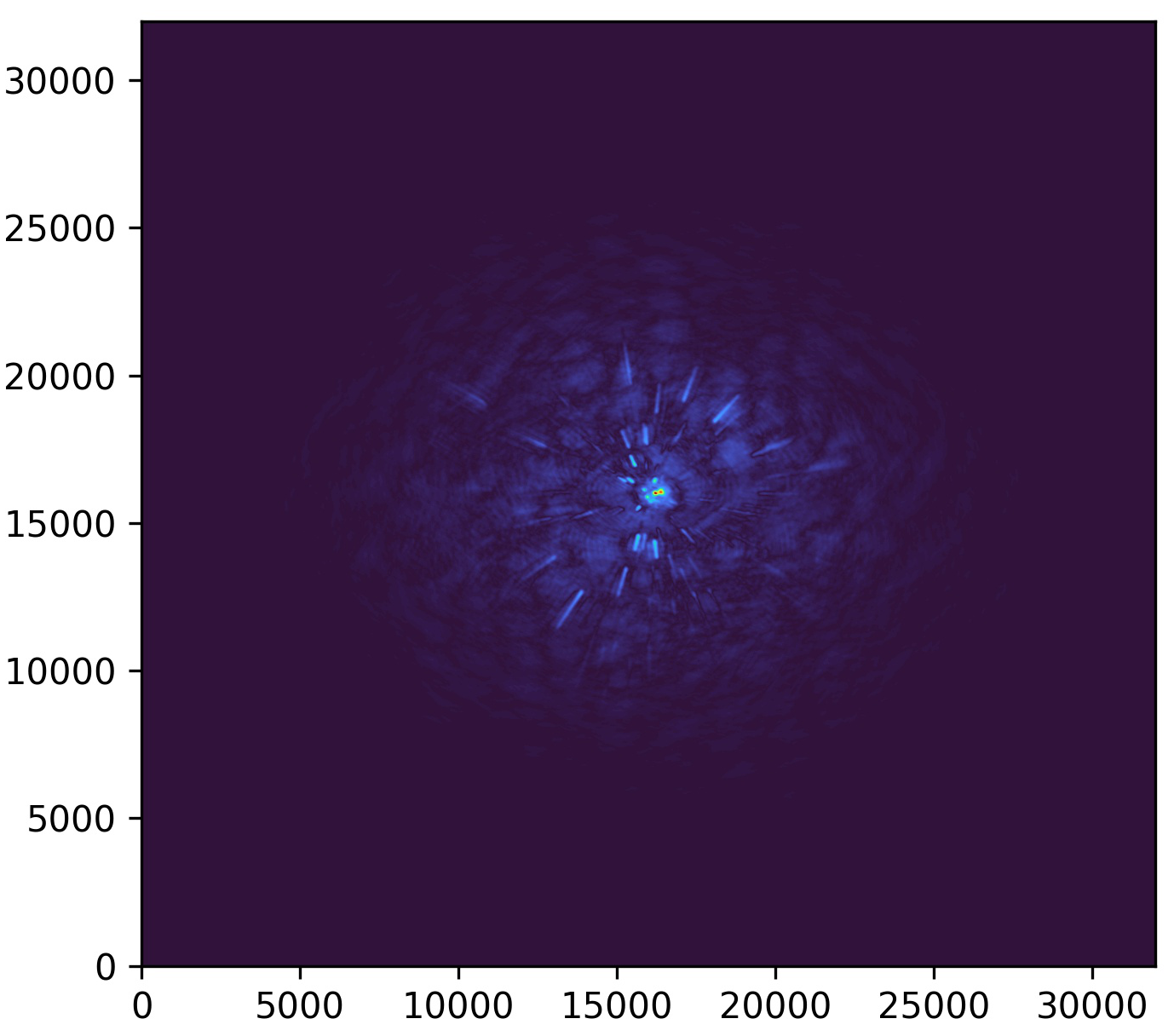}
\caption{Result of the w-stacking procedure applied to the Large datasets producing a 4096$\times$4096 pixels (left panel) and a 32768$\times$32768 pixels (right panel) image. The different mesh sizes reflect on the different fields of view of the two images. Smearing of outermost sources is due to the artificial combination of different bandwidths.}
\label{fig:maps}
\end{figure*}

The test requires about 2.5 TB of memory, mapping to 128 nodes of the M100 system. 
Both pure MPI and GPU set-up have been run, the former using 4096 MPI tasks (32 cores per node), the latter 512 (4 GPUs per node). In both cases the time to solution resulted to be between 2750 and 3150 seconds. In Figure \ref{fig:maps}, the 32768$\times$32768 pixels image is compared to a 4096$\times$4096 pixels one. The latter is the minimum image size which ensures the nominal observation resolution ($\sim$6 arcsec) to be sampled by at least 2 pixels. The increased resolution of the computational mesh of the former reflects on its much larger field of view, the 4096$\times$4096 field corresponding to a small square tile in the centre of the 32768$\times$32768 image. The outermost sources are smeared due to the artificial combination of different bands, that would normally be treated separately and then properly merged, but that was required by our test to fulfil the computational requirements. 

It is interesting to notice that the same memory requirements of the Large test would be needed in order to process a 262144$\times$262144$\times$1 mesh. This means that a 262144$\times$262144 pixels image can be generated with the above set-up.

\section{Conclusions}

In this paper, we have explored possible effective strategies toward the support of large radio-interferometric datasets, as those expected from the SKA radiotelescope, by exploiting state-of-the-art High Performance Computing solutions. We have focused on the imaging algorithm, specifically addressing the gridding, FFT transform and w-correction steps, in terms of both memory request and computing time. 

Our w-stacking gridder has been developed to demonstrate high performance and scalability on modern accelerated, multi-processors devices, addressing also portability and maintainability, in order to facilitate the usage of the same software on different computing systems in a time scale much longer than that expected for a typical HPC technology, maximising its usability and impact. 

Our main achievements can be summarised as follows:
\begin{itemize}
    \item we have enabled a prototype w-stacking gridder to the usage of heterogeneous HPC solutions, combining parallel computing with accelerators.
    \item The resulting code supports dataset whose size is limited only by the physical memory of the available computing units, evenly distributing data among distributed memories. Our proof of concepts run could perform the  w-stacking of a 102 GB input dataset on a 32768$\times$32768$\times$64 cells mesh {\it in memory} (overall memory request of about 2.5 TB) in less than one hour. 
    \item The scalability of the code (both strong and weak) ensures that it can be efficiently used on large HPC configurations, involving hundreds or even thousands of cooperating computing units.
    \item The usage the GPUs reduces the time to solution thanks to their outstanding performance, hence allowing using a smaller number of MPI tasks compared to a pure MPI set-up, alleviating communication related issues. 
    \item Communication overheads can have a strong impact on the code performance when large images are generated using hundreds or thousands of computing units. However, the advantage of performing the full calculation in memory maintains the time to solution within reasonable limits (less than 1 hour for our biggest case).
    \item The code is easily portable on different systems, being written in standard C with some limited extensions to C++ and avoiding any tuning to a specific architecture. It can be compiled with a GCC GNU compiler and, in case NVIDIA GPUs are used, by the NVCC compiler. The only dependency, besides MPI, is from the FFTW3 library, which, however, can be easily installed on any computing system. 
    \item The usage of the offload support of OpenMP allowed achieving the porting to the GPU also adopting simple directives, in principle portable to different accelerated architectures, without (in this case) any performance penalty compared to CUDA. However, offload directives are fully supported only by GCC, version $>$ 9, or NVCC version $>$ 10.
    \item The usage of HPC systems enables the processing of data and images of unprecedented size, particularly important for very large datasets and/or low frequency and very large baselines arrays.
\end{itemize}
This work represents only the first step toward a full exploitation of ultimate HPC solutions in radio astronomy. Additional steps are already taken in the direction of optimising the usage of the GPUs (e.g. exploiting the CUDA Fast Fourier Transform library, cuFFT, whose multi-node version has been released at the beginning of 2022), which proved to deliver promising results in terms of performance, and extending the code to the exploitation of other kind of accelerators, in particular FPGA architectures. A critical step will be also the improvement of the performance related to communication both exploring novel patterns to minimise data exchange but also to acquire an in-depth comprehension of the MPI data transfer and synchronisation mechanisms on a given interconnect, in order to exploit it at full efficiency. Finally, we will extend our study to include additional features toward the support of the full imaging procedure. More specifically we will introduce   convolutional kernels commonly used in radio astronomy, as, for instance, the prolate spheroidal wave function or the Kaiser-Bessel window function \citep{jackson_91} and we will extend our parallel approach to the degridding procedure (transforming from the image to the visibility space). The same parallelization strategy is expected to be effective also for such inverse process, although a  specific implementation will be required by the distribution of the visibilities on the cell back to the single measurements in the $(u,v,w)$ space. Furthermore, we will support the full treatment of multi-wavelength observations and of different polarisation.

\section*{Data Availability}
\label{data}
The data used for this work have been produced within the LOFAR project LC14\_018 and will be available according to the standard LOFAR policy.

\section*{Acknowledgements}

The authors want to thank Francesco De Gasperin, Luca Bruno, Luca tornatore and Gianfranco Brunetti for the valuable discussions, advice, remarks and suggestions that helped improving the contents of the paper. The HPC tests and benchmarks this work is based on, have been produced on the Marconi100 Supercomputer at CINECA (Bologna, Italy) in the framework of the ISCRA programme, IscrC\_TRACRE project, and on the Juwels Booster Module at the J\"ulich Supercompting Centre (Germany) thanks to the support by the European Union's Horizon 2020 Research and Innovation Programme under the EuroEXA project (Grant Agreement No. 754337). The data to perform all the tests have been kindly provided by the LOFAR project LC14\_018, PI Franco Vazza.

\bibliographystyle{mnras}
\bibliography{franco,franco2,bib_add}

\bsp	
\label{lastpage}
\end{document}